\documentclass[preprint]{elsarticle}
\journal{International Journal of Critical Infrastructure Protection}
\usepackage[margin=1in]{geometry}
\usepackage{lineno}
\usepackage{url}
\usepackage{xurl}
\usepackage{arydshln}
\usepackage{subfig}
\usepackage{bbm}
\usepackage{bm}
\usepackage{longtable}
\usepackage{amsmath,amssymb,amsfonts}
\usepackage{tensor}
\usepackage{esint}
\usepackage{algpseudocode}
\usepackage{graphicx}
\usepackage{textcomp}
\usepackage{adjustbox}
\usepackage{mdframed}
\usepackage{xcolor}
\usepackage{mathtools, nccmath, cuted}
\usepackage{siunitx}

\usepackage{comment}
\usepackage{makecell}
\usepackage{algorithm}
\usepackage{gensymb}
\usepackage[normalem]{ulem}
\usepackage{colortbl}
\usepackage{enumitem}
\definecolor{Gray}{gray}{0.925}
\usepackage[textsize=tiny]{todonotes} 
\usepackage{tabularx}

\graphicspath{ {figures/} }
\def\BibTeX{{\rm B\kern-.05em{\sc i\kern-.025em b}\kern-.08em
    T\kern-.1667em\lower.7ex\hbox{E}\kern-.125emX}}

\usepackage{pgf}
\usepackage{pgfplots}

\usetikzlibrary{shapes,arrows,calc}
\usetikzlibrary{arrows.meta}
\usetikzlibrary{decorations.pathreplacing}
\usetikzlibrary{shapes.geometric}
\usetikzlibrary{shadows}

\usepackage{pgfplotstable}
\pgfplotsset{compat=1.14}
\usepackage{listings}
\usepackage{array,colortbl,multirow,multicol,booktabs,ctable}

\definecolor{delim}{RGB}{0,0,0}
\definecolor{numb}{RGB}{0, 0, 0}
\definecolor{string}{rgb}{0.54,0.68,0.08}
\usepackage{float}

\lstdefinelanguage{json}{
    numbers=left,
    numberstyle=\small,
    frame=single,
    rulecolor=\color{black},
    showspaces=false,
    showtabs=false,
    breaklines=true,
    postbreak=\raisebox{0ex}[0ex][0ex]{\ensuremath{\color{gray}\hookrightarrow\space}},
    breakatwhitespace=true,
    basicstyle=\ttfamily\small,
    upquote=true,
    morestring=[b]",
    stringstyle=\color{string},
    literate=
     *{0}{{{\color{numb}0}}}{1}
      {1}{{{\color{numb}1}}}{1}
      {2}{{{\color{numb}2}}}{1}
      {3}{{{\color{numb}3}}}{1}
      {4}{{{\color{numb}4}}}{1}
      {5}{{{\color{numb}5}}}{1}
      {6}{{{\color{numb}6}}}{1}
      {7}{{{\color{numb}7}}}{1}
      {8}{{{\color{numb}8}}}{1}
      {9}{{{\color{numb}9}}}{1}
      {\{}{{{\color{delim}{\{}}}}{1}
      {\}}{{{\color{delim}{\}}}}}{1}
      {[}{{{\color{delim}{[}}}}{1}
      {]}{{{\color{delim}{]}}}}{1},
}

\definecolor{blueLine}{RGB}{57,106,177}
\definecolor{blueFill}{RGB}{114,147,203}
\definecolor{redLine}{RGB}{204,37,41}
\definecolor{greenline}{RGB}{0,250,0}
\definecolor{blackLine}{RGB}{0,0,0}
\definecolor{goldLine}{RGB}{160,82,45}
\definecolor{goodGreen}{RGB}{213,232,212}
\definecolor{goodGreenBorder}{RGB}{150,192,129}
\definecolor{goodBlue}{RGB}{218,232,252}
\definecolor{goodBlueBorder}{RGB}{144,170,207}
\definecolor{goodPink}{RGB}{248,206,204}
\definecolor{goodPinkBorder}{RGB}{200,114,111}
\definecolor{airforceblue}{rgb}{0.36, 0.54, 0.66}
\definecolor{aquamarine}{rgb}{135, 206, 255}
\definecolor{deepskyblue}{rgb}{0.0, 0.75, 1.0}
\definecolor{persianblue}{rgb}{0.11, 0.22, 0.73}
\definecolor{aliceblue}{rgb}{0.94, 0.97, 1.0}

\usepackage[font=footnotesize]{caption}
\usepackage{hyperref}
\makeatletter
\def\ps@pprintTitle{%
  \let\@oddhead\@empty\let\@evenhead\@empty
  \def\@oddfoot{\footnotesize\itshape Preprint submitted to International Journal of Critical Infrastructure Protection\hfill\today}%
  \let\@evenfoot\@oddfoot}
\makeatother

\begin{document}
\begin{frontmatter}

\title{Cyber security of OT networks: A tutorial, survey of attacks and overview of current state of defense tools, protocols, \& challenges}

\author[add1]{Harsh Vardhan\corref{corrauth}\fnref{eq}}
\ead{harsh.vardhan@vanderbilt.edu}
\author[add2]{Sarthak Kapoor\fnref{eq}}
\author[add3]{Sumit Kumar\fnref{eq}}
\author[add1]{Daniel Balasubramanian}
\author[add1]{Sandeep Neema}
\cortext[corrauth]{Corresponding author}
\fntext[eq]{These authors contributed equally to this work.}

\address[add1]{Vanderbilt University, USA}
\address[add2]{Engineering at Amazon, USA}
\address[add3]{Georgia State University, USA}
\date{}

\begin{abstract}
The convergence of Operational Technology (OT) and Information Technology (IT) under Industry~4.0 has widened the cyber-attack surface of critical infrastructure across manufacturing, energy, transportation, water, and healthcare. This survey synthesizes OT/IT cybersecurity along four axes. First, we taxonomize \emph{attack vectors} that traverse the IT--OT boundary, separating IT-side initial access (phishing, exploits, supply-chain compromise, exposed remote access) from OT-side propagation and impact (insecure protocols, weak authentication, firmware tampering, control-logic manipulation). Second, we review \emph{defensive technologies}---signature-based intrusion detection, AI/ML anomaly detection, Zero Trust Architecture, blockchain-based event logging, digital twins, and OT-aware Security Operations Centers---and identify remaining \emph{gaps}: OT-specific patch management, dataset scarcity for ML, IoMT segmentation, and the absence of consistent resilience metrics. Third, we compile a cross-validated \emph{historical record} of 69 high-impact incidents spanning 2010--2025, from Stuxnet to Jaguar Land Rover, and quantify their \emph{commercial effects} sector by sector using figures sourced from SEC filings, government post-incident reviews, and primary regulatory disclosures. Fourth, we map the \emph{regulatory landscape}: NIST Cybersecurity Framework~2.0 and SP~800-82~Rev.~3, IEC~62443, the EU NIS2 Directive, DORA, the Cyber Resilience Act, NERC~CIP, and healthcare-specific regimes (IEC~80001-1, FDA, NIST~SP~1800-8). A sectoral deep-dive on healthcare illustrates the IT--OT convergence threat model under high-consequence conditions. The result is a single, source-traceable reference on where OT/IT cybersecurity stands, what the historical record costs defenders who lag, and where investment yields the highest marginal return on resilience.
\end{abstract}


\end{frontmatter}

\section{Introduction}
\label{sec:introduction}

Operational Technology (OT) networks govern the physical processes that underpin modern society---electric power generation and distribution, oil and gas transport, water treatment, discrete and continuous manufacturing, rail and port logistics, and increasingly hospital facilities and biomedical production. These networks are built from Supervisory Control and Data Acquisition (SCADA) systems, Programmable Logic Controllers (PLCs), Distributed Control Systems (DCS), Remote Terminal Units (RTUs), Human--Machine Interfaces (HMIs), and field instrumentation that close real-time control loops over deterministic protocols.

For most of their history, OT networks were physically and logically isolated from corporate Information Technology (IT) environments. That isolation has eroded. Industry~4.0, predictive-maintenance analytics, cloud-based historians, remote engineering access, and the proliferation of Industrial Internet of Things (IIoT) and Internet of Medical Things (IoMT) endpoints have driven deep IT--OT convergence \cite{negi2024towards, knapp2015securing, zaid2024emerging}. The same convergence that yields operational visibility and analytic insight also exposes legacy controllers---designed for availability and determinism, not adversarial resilience---to a much larger attack surface. The consequences are no longer limited to data loss. Stuxnet (2010) physically damaged centrifuges \cite{Langner, kushner2013real}; Industroyer (2016) tripped circuit breakers at a transmission substation near Kyiv \cite{kozak2023industroyer}; Triton (2017) reached safety instrumented systems in a Saudi petrochemical plant \cite{mekdad2021threat}; Colonial Pipeline (2021) halted a pipeline transporting approximately 45\% of fuel supplied to the U.S.\ East Coast for six days \cite{cisa2023colonial, guardian2021colonial}; and INCONTROLLER/PIPEDREAM (2022) revealed a modular, vendor-agnostic toolkit purpose-built for ICS disruption \cite{incontroller2022, kalinaki2025ransomware, daniel2024emerging}.

\subsection{Scope and Research Questions}
This survey addresses cybersecurity at the IT--OT interface and within OT proper. We scope this survey around four questions a practitioner or researcher entering the field must answer:

\begin{enumerate}
    \item \textbf{RQ1 (Vectors).} Which attack vectors traverse the IT--OT boundary, and how do attacker techniques on the IT side translate into impact on OT assets?
    \item \textbf{RQ2 (Tools).} What detective, preventive, and architectural controls are available, and how do they compare on accuracy, latency, adaptability, and operational fit for OT?
    \item \textbf{RQ3 (Gaps).} Where do current tools, datasets, standards, and operational practices fall short?
    \item \textbf{RQ4 (Impact).} What does the historical incident record tell us about commercial and societal effects, and which sectors and cost components dominate?
\end{enumerate}

\textbf{Out of scope:} pure IT security topics with no OT-specific implication (general-purpose enterprise endpoint protection, web-application security), physical security and insider threats unrelated to cyber compromise, and detailed cryptanalysis of industrial protocols.

\subsection{Contributions}
This work makes the following contributions:

\begin{enumerate}
    \item A consolidated \textbf{attack-vector taxonomy} that links IT-side initial-access techniques to OT-side propagation and impact, with a representative vector-to-incident mapping covering USB, phishing, supply-chain compromise, internet-exposed PLCs, helpdesk social engineering, and wartime wiper operations (\S\ref{sec:ga_sbo}).
    \item A \textbf{comparative review of defensive tools}---signature- and rule-based IDS, AI/ML anomaly detection, Zero Trust, blockchain logging, digital twins---against published evaluations, with an explicit \textbf{gap analysis} that identifies where research and engineering investment is most needed (\S\ref{sec:tools_gaps}).
    \item A source-traceable \textbf{historical incident record} from 2010--2025 (69 cross-validated incidents), enriched with per-sector cost and downtime data drawn from primary disclosures, to quantify the commercial impact of OT cyber incidents (\S\ref{sec:history_impact}).
    \item A discussion of \textbf{open problems} at the intersection of OT determinism, ML adaptivity, and standards-based assurance (\S\ref{sec:conclusion}).
\end{enumerate}

\subsection{Survey Methodology}
\label{subsec:methodology}
The corpus underlying this survey was assembled in three passes. \emph{First}, we performed keyword searches in IEEE Xplore, ACM Digital Library, ScienceDirect, SpringerLink, and Google Scholar using combinations of the terms \textit{operational technology}, \textit{industrial control system}, \textit{SCADA}, \textit{PLC}, \textit{cybersecurity}, \textit{intrusion detection}, \textit{IT--OT convergence}, \textit{Industry~4.0 security}, and \textit{cyber resilience}. The academic-literature search was restricted to publications from 2014 through 2025, with primary incident-analysis sources retained back to 2010 where they remain canonical (e.g., Stuxnet analyses).

\emph{Second}, we supplemented academic sources with primary incident reports and threat intelligence from CISA advisories, MITRE ATT\&CK for ICS, ENISA, vendor reports (Dragos, Claroty, Fortinet, Kaspersky, IBM X-Force, Mandiant), and standards documents (NIST CSF~2.0, NIST~SP~800-82~Rev.~3, IEC~62443 series, IEC~80001-1, EU NIS2, DORA, EU CRA).

\emph{Third}, we performed forward and backward citation tracing on a seed set of high-impact survey and incident-analysis papers to capture work missed by keyword search. We excluded sources without primary attribution for empirical claims. Numerical estimates retained in the paper are attributed to their issuing organization and year, and clearly marked as \emph{tail-event scenario estimates} or \emph{survey-based} where the underlying methodology is itself a tail-event model or industry survey rather than an instrumented measurement.

\subsection{Paper Organization}
Section~\ref{sec:ot} provides background on OT components (SCADA, DCS, PLCs, RTUs) and the protocols that interconnect them. Section~\ref{sec:ga_sbo} taxonomizes attack vectors at the IT--OT boundary. Section~\ref{sec:tools_gaps} surveys defensive tools, compares their performance characteristics, and identifies open gaps. Section~\ref{sec:history_impact} compiles a cross-validated historical incident record from 2010 through 2025 with sector-by-sector quantitative analysis of commercial impact and a deep-dive on healthcare. Section~\ref{sec:reg_frameworks} reviews the U.S., European, and international regulatory frameworks governing OT cybersecurity (NIST CSF~2.0, NIST~SP~800-82, IEC~62443, NIS2, DORA, the EU Cyber Resilience Act, NERC CIP, and healthcare-specific regimes), referencing the incident and impact data of the prior section. Section~\ref{sec:related_work} positions our work against prior surveys. Section~\ref{sec:conclusion} concludes with open problems.

\section{Background: OT Systems and IT--OT Convergence}
\label{sec:ot}

This section establishes the vocabulary used throughout the survey. We summarize the building blocks of OT environments---SCADA, DCS, PLCs, RTUs, communication protocols---and articulate why their convergence with IT changes the threat model.

\subsection{Industrial Control Systems: SCADA, DCS, PLC, RTU}

\textbf{SCADA (Supervisory Control and Data Acquisition)} systems provide centralized monitoring and supervisory control of geographically distributed processes. A SCADA system integrates a sensing layer (field instruments), a control layer (programmable controllers), and a supervisory layer (master station, HMI, historian). SCADA is typical of power transmission, oil and gas pipelines, water distribution, and rail signaling \cite{waqas2024smart}.

\textbf{Distributed Control Systems (DCS)} decentralize control across cooperating controllers within a localized facility. DCS is typical of chemical, petrochemical, and power-generation plants, where complex closed-loop control of many tightly coupled variables is required. The boundary between SCADA and DCS has blurred as both adopt Ethernet-based networking and shared HMI conventions, but the canonical distinction remains: SCADA emphasizes wide-area supervisory data acquisition; DCS emphasizes localized continuous control with heavy inter-controller communication.

\textbf{Programmable Logic Controllers (PLCs)} are deterministic real-time controllers that scan inputs, execute control logic, and update outputs on millisecond cycles. They are typically programmed under the IEC~61131-3 standard \cite{iec61131_3} using Ladder Logic, Function Block Diagram, Structured Text, Sequential Function Chart, or Instruction List. PLCs are the workhorses of factory floors, water utilities, building management, and increasingly hospital facility OT.

\textbf{Remote Terminal Units (RTUs)} are microprocessor-based devices that interface field sensors and actuators with the supervisory layer over long-distance media (cellular, radio, satellite) or wired Ethernet links \cite{misbahuddin2010fault}. They perform local data acquisition, digitization, and limited control, and report to a SCADA master.

\textbf{Human--Machine Interfaces (HMIs)} present operators with process visualization, alarm management, and command issuance. \textbf{Historians} archive time-stamped process data for trend analysis, regulatory reporting, and predictive maintenance.

\begin{figure}[ht!]
    \centering
   \includegraphics[trim={0.8cm 0.5cm 0.5cm 0.1cm},clip,width=0.9\linewidth]{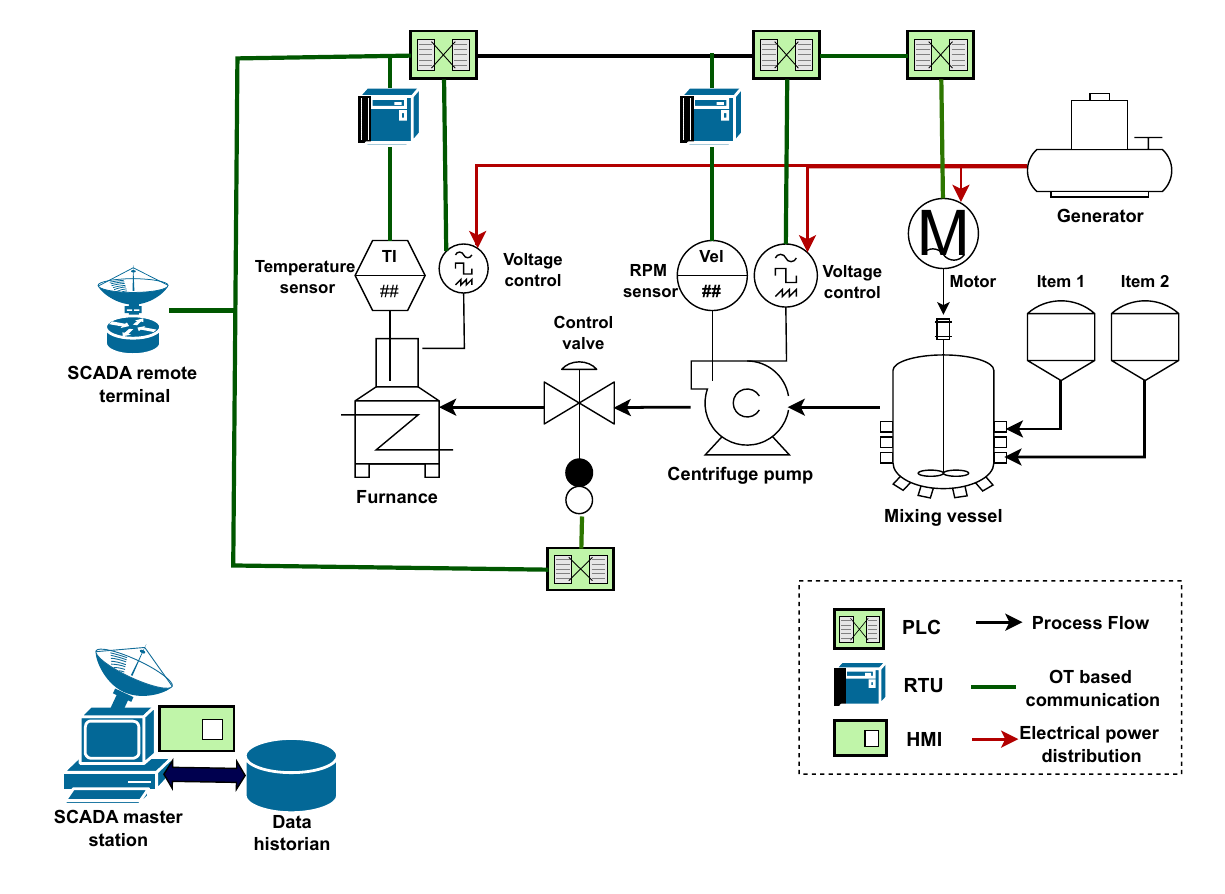}
    \caption{An example mixing-process plant with an OT network: PLC-based controllers drive field actuators (motors, valves, heaters); RTUs aggregate sensor data; a SCADA master with HMI provides supervisory visibility and a historian for archival. Power distribution is shown in red; OT communication flows in green.}
    \label{fig:scada}
\end{figure}

\subsection{Communication Protocols}
OT networks use a heterogeneous mix of communication protocols, many of which predate modern security thinking. Table~\ref{tab:protocols} summarizes the most common protocols and their typical applications. Several of these (notably Modbus, DNP3 unencrypted, EtherNet/IP, Profinet) lack built-in authentication or encryption, a limitation we revisit when discussing OT-side attack vectors in Section~\ref{sec:ga_sbo}.

\begin{table}[h!]
\centering
\footnotesize
\begin{tabular}{|m{3.0cm}|m{5.0cm}|m{6.5cm}|}
\hline
\textbf{Protocol} & \textbf{Description} & \textbf{Common Applications} \\ \hline
\textbf{Modbus (RTU/ASCII/TCP)} & Open master/slave protocol; simple message structure; no native authentication or encryption. & Process automation, SCADA, sensor-to-PLC communication. \\ \hline
\textbf{Profibus (DP/PA)} & High-speed fieldbus with extensive diagnostics; DP for discrete I/O, PA for process automation. & Manufacturing, process automation. \\ \hline
\textbf{Profinet} & Ethernet-based; real-time, IT-integrable. & Industrial automation requiring high-speed deterministic control. \\ \hline
\textbf{EtherNet/IP} & Industrial protocol over standard Ethernet; widely used by Rockwell/Allen-Bradley devices. & Factory automation, robotics, assembly lines. \\ \hline
\textbf{DNP3} & Designed for SCADA reliability; supports time-stamped data. Secure variant (DNP3-SA) adds authentication. & Electric utilities, water/wastewater, oil and gas. \\ \hline
\textbf{IEC~60870-5-104} & Telecontrol over TCP/IP; common in European power utilities. & Power transmission and distribution. \\ \hline
\textbf{IEC~61850} & Object-oriented modeling for substation automation; GOOSE and Sampled Values for low-latency communication. & Electric substations, smart grid. \\ \hline
\textbf{OPC~UA} & Modern, platform-independent protocol with built-in authentication and encryption. & Cross-vendor interoperability; bridge between OT and IT. \\ \hline
\textbf{HART} & Combines analog (4--20\,mA) with digital signals over the same wiring. & Process-industry instrumentation. \\ \hline
\textbf{BACnet} & Open building-automation protocol. & HVAC, lighting, fire detection in commercial buildings. \\ \hline
\textbf{CANopen / DeviceNet} & CAN-based; real-time, low-cost. & Automotive, embedded systems, sensor/actuator networks. \\ \hline
\end{tabular}
\caption{Common communication protocols in OT environments. OPC~UA and DNP3-SA are the principal options when authentication and encryption are required.}
\label{tab:protocols}
\end{table}

\subsection{IT--OT Convergence and Why It Matters}

Three trends have driven IT--OT convergence: (i)~\emph{remote operations} requiring access to plant data from corporate networks and the cloud; (ii)~\emph{predictive analytics and AI/ML} consuming historian data at scale; and (iii)~\emph{IIoT proliferation}, with low-cost sensors and edge devices sharing the same Ethernet and IP networks as engineering workstations and PLCs. The architectural consequence---a zoned, gateway-mediated boundary between corporate IT and plant-floor OT---is illustrated in Section~\ref{sec:ga_sbo} (Figure~\ref{fig:itot}), where it serves as the canvas for the attack-vector discussion.

The security implications of this convergence differ from those of pure IT environments in four ways. (1)~\emph{Availability dominates confidentiality}: a control loop that pauses for a security update can produce off-spec product or unsafe states. (2)~\emph{Long lifecycles}: PLCs and DCS controllers commonly remain in production for 15--25 years, well beyond their vendor-supported window---a characteristic recurrently noted in NIST SP~800-82~Rev.~3\cite{nist80082rev3} and Dragos year-in-review reporting \cite{o2024fiscal}. (3)~\emph{Real-time determinism}: cryptographic overhead and detection latency must respect cycle-time budgets. (4)~\emph{Cyber-physical impact}: a successful attack can damage equipment, harm personnel, or interrupt critical services---consequences that have no analog in conventional IT breaches. These four properties shape the attack vectors and defensive options discussed in the remainder of the survey.

\section{Attack Vectors}
\label{sec:ga_sbo}

This section taxonomizes the attack vectors that threaten OT environments. We distinguish two phases of an end-to-end campaign: (i)~\emph{IT-side initial access and lateral movement}, where adversaries exploit conventional enterprise weaknesses to reach the IT--OT boundary, and (ii)~\emph{OT-side propagation and impact}, where attackers exploit the unique properties of industrial protocols, controllers, and engineering workflows to manipulate physical processes. The historical incidents that exercised these vectors in practice are catalogued in Section~\ref{sec:history_impact}.

\begin{figure}[ht!]
    \centering
   \includegraphics[trim={0.8cm 0.5cm 0.5cm 0.1cm},clip,width=0.92\linewidth]{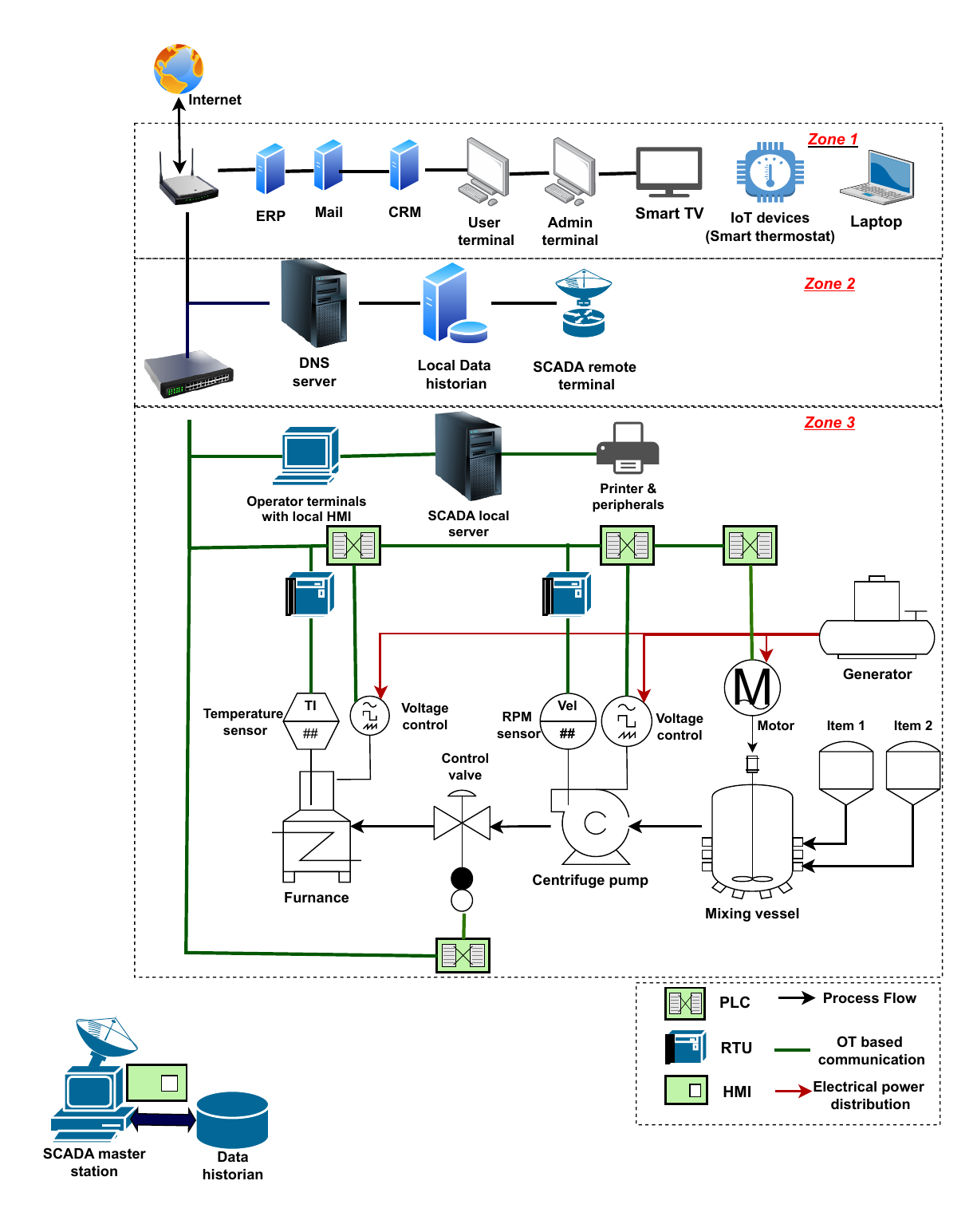}
    \caption{Representative IT--OT architecture used as the reference model for the attack-vector discussion. Zone~1 is the corporate IT network, Zone~2 is the IT--OT bridge (industrial DMZ with historian replica, jump server, and unidirectional gateway), and Zone~3 is the OT network containing supervisory, control, and field layers.}
    \label{fig:itot}
\end{figure}

Figure~\ref{fig:itot} provides the architectural canvas for this discussion. Zone~1 is the corporate IT network, where most initial-access vectors are first realized. Zone~2 is the IT--OT bridge (an industrial DMZ that typically hosts a historian replica, jump server, DNS, and a unidirectional gateway), where the majority of segmentation and inspection controls operate. Zone~3 is the OT network proper, where supervisory, control, and field layers interact with physical process equipment. Each attack vector below is described in terms of the zone in which it is realized and the path it takes across this boundary.

\subsection{IT-Side Vectors: Reaching the OT Boundary}
\label{subsec:itvectors}

Most documented OT compromises begin in the IT environment and pivot inward. Five vector families dominate.

\textbf{Phishing and social engineering.} Spear-phishing emails targeting engineers, plant operators, or vendor personnel remain the single most common initial-access technique. Attackers exploit the trust placed in vendor support communications and the tendency for engineering workstations to dual-home onto IT and OT networks. The 2015 BlackEnergy intrusions into Ukrainian distribution operators began with macro-laden Word attachments delivered to dispatch staff, with subsequent multi-month dwell before the December 2015 outage \cite{ukraine2015cisa, ukraine2015sans}.

\textbf{Vulnerability exploitation.} Unpatched Internet-facing services (VPN gateways, Citrix appliances, Exchange servers, web portals) and outdated workstations provide initial footholds. Once inside, attackers exploit Active Directory weaknesses, weak service-account credentials, and cached privileges to move laterally toward engineering systems. CVE-driven exploitation of remote-access infrastructure was the dominant initial-access vector in OT incidents catalogued by Dragos in 2023--2024.

\textbf{Supply-chain compromise.} Compromised vendor software updates, malicious firmware images, and trojanized hardware introduce vulnerabilities that bypass perimeter controls. The SolarWinds Orion compromise (2020) \cite{martinez2021software} and the Kaseya VSA ransomware campaign (2021) \cite{robinson2022new} demonstrated that trusted update channels can deliver attacker payloads at scale; the Kaseya campaign in particular reached OT-operating customers via managed-service-provider relationships.

\textbf{Insecure remote access.} Remote engineering and vendor-maintenance access is operationally indispensable but historically poorly secured. Weak or shared credentials on jump servers, unmanaged third-party VPN tunnels, and exposed Remote Desktop Protocol (RDP) endpoints create direct paths into engineering networks. EKANS/Snake (2019--2020) \cite{dragos2020ekans} and multiple ransomware incidents in manufacturing during 2020--2022 entered through internet-exposed RDP.

\textbf{Removable media and engineering laptops.} Where strict network segmentation prevents network-based ingress, USB drives and engineering laptops carried between IT and OT zones become the bridging mechanism. Stuxnet (2010) used USB removable media as the bridging mechanism. The malware exploited the Windows Shell LNK-parsing vulnerability (CVE-2010-2568, MS10-046), which executed shortcut payloads on icon display in Windows Explorer---not the Autorun mechanism, which could be disabled without preventing the exploit---to propagate across the air-gapped Natanz network.

\subsection{OT-Side Vectors: Propagation and Impact}
\label{subsec:otvectors}

Once an adversary obtains a foothold near the IT--OT boundary, the properties of OT itself enable further propagation and physical-process impact. Six vector families are documented in incident reports and academic literature.

\textbf{Insecure communication protocols.} Modbus, DNP3 (without secure-authentication extensions), EtherNet/IP, and many vendor-proprietary protocols lack message authentication, encryption, and integrity checks. This enables \emph{eavesdropping} on process variables and setpoints, \emph{man-in-the-middle} modification of commands between the engineering workstation and the PLC, and \emph{replay} of legitimate commands to induce arbitrary actuator behavior \cite{s7threats}. OPC~UA, DNP3-SA, and TLS-wrapped variants address these gaps but are unevenly deployed in installed plant.

\textbf{Weak authentication and credential management.} Default vendor credentials, hard-coded service accounts, shared engineering passwords, and absence of multi-factor authentication are pervasive in deployed PLCs and HMIs. Many controllers still expose unauthenticated programming interfaces over the network. Brute-force and credential-stuffing against engineering workstations and PLC web servers are routine.

\textbf{Firmware vulnerabilities and unsigned updates.} A subset of PLCs accepts firmware updates without verifying authenticity, which permits attackers to inject persistent malicious firmware. Vendor-introduced backdoors and unauthenticated programming interfaces have been documented in Siemens S7 controllers across multiple disclosure rounds, beginning with Beresford's 2011 Black Hat USA work on the S7 series (covering S7-300/400 and S7-1200, including a hard-coded password and ISO-TSAP authentication bypass) \cite{beresford2011siemens} and continuing through later disclosures of unauthenticated update paths in the S7-1200/S7-1500 series \cite{enlyze, s7threats}.

\textbf{Patch-management and lifecycle gaps.} OT systems are patched infrequently because of operational continuity requirements, vendor revalidation cycles, and the absence of OT-aware patch-management tooling. Legacy controllers (Windows~XP-based engineering stations, end-of-life PLC firmware) often remain in service after vendor support ends, leaving known vulnerabilities permanently unaddressed.

\textbf{Insufficient access control and audit.} PLCs frequently lack fine-grained role-based access control and produce minimal logs. Attackers can issue critical commands without leaving an audit trail, and forensics after an incident is limited to network captures and SCADA-side records. Mature deployments compensate by mirroring and analyzing OT traffic on dedicated taps.

\textbf{Side-channel and physical-layer attacks.} Power-analysis, electromagnetic-emanation, and timing attacks on PLCs and cryptographic accelerators are documented in the academic literature and demonstrated in lab settings. They are not yet a mainstream operational threat but are an active research area, particularly for safety-critical and defense-related deployments.

\subsection{End-to-End Attack Paths}
\label{subsec:endtoend}

Mapping IT-side and OT-side vectors onto the cyber kill chain produces the end-to-end paths summarized in Table~\ref{tab:vector_taxonomy}. Each row tracks an attack from the IT-side initial-access technique through the OT-side mechanism that produces physical-process impact, with a representative incident as evidence.

\begin{table}[h!]
\centering
\footnotesize
\caption{Attack-vector taxonomy mapping IT-side initial access to OT-side impact, with representative incidents (detailed in Section~\ref{sec:history_impact}).}
\label{tab:vector_taxonomy}
\renewcommand{\arraystretch}{1.2}
\begin{tabular}{|m{3.0cm}|m{4.0cm}|m{4.5cm}|m{3.2cm}|}
\hline
\textbf{IT-side vector} & \textbf{Pivot mechanism} & \textbf{OT-side impact mechanism} & \textbf{Representative incident} \\ \hline
USB / removable media & USB insertion triggers .LNK shell-parsing exploit (CVE-2010-2568) on engineering workstation & Malicious payload reprograms PLC logic; rootkit hides altered control behavior & Stuxnet (2010) \cite{Langner, kushner2013real} \\ \hline
Spear phishing of utility staff & Credential theft; lateral movement to ICS network & Direct manipulation of ICS protocols (IEC~60870-5-104, IEC~61850) to open breakers & Ukraine 2015 \cite{ukraine2015cisa, ukraine2015sans}; Industroyer 2016 \cite{kozak2023industroyer} \\ \hline
Compromise of engineering workstation & Vendor protocol abuse; SIS-controller reprogramming & Modification of safety-instrumented-system logic & Triton/Trisis (2017) \cite{mekdad2021threat, trisis_drago} \\ \hline
Phishing or RDP brute force & Lateral movement using legitimate admin tools (PowerShell, PsExec) & ICS-process termination + file encryption on engineering and historian hosts & EKANS/Snake (2019--2020) \cite{dragos2020ekans}\\ \hline
Exploit of public-facing application & Network reconnaissance; PLC fingerprinting & Worm-style scanning and infection of additional PLCs over industrial protocols & PLC-Blaster (2016) \cite{spenneberg2016plcblaster} \\ \hline
Vendor software supply-chain compromise & Trusted update delivers attacker payload & Modular toolkit issues vendor-specific commands across multiple PLC families & INCONTROLLER / PIPEDREAM (2022) \cite{incontroller2022} \\ \hline
Internet-exposed PLC + default credentials & Direct PLC programming over the Internet & HMI defacement; manual fallback at small utilities & CyberAv3ngers / Aliquippa (2023) \cite{cyberavengers2023}; Iran-affiliated PLC campaign (2025--26) \cite{iranplc2026} \\ \hline
Helpdesk social engineering & Active Directory takeover via support-desk ticket & IT-side encryption forcing operational shutdown across retailers and manufacturers & M\&S/Co-op/JLR (Scattered Spider, 2025) \cite{ms_coop_2025} \\ \hline
Wartime VPN/management-plane abuse & Wiper deployed against modems or sensors & Loss of remote SCADA control or destructive bricking of OT-side gateways & AcidRain/KA-SAT (2022) \cite{acidrain2022}; Fuxnet (2024) \cite{fuxnet2024} \\ \hline
Exposed remote access (VPN/RDP) & Credential reuse; ransomware deployment & Operational disruption via IT-side encryption forcing OT shutdown & Colonial Pipeline (2021) \cite{cisa2023colonial, guardian2021colonial}; CDK Global (2024) \cite{cdkglobal2024} \\ \hline
\end{tabular}
\end{table}

The taxonomy makes two observations explicit. \emph{First}, the IT-side vector and the OT-side impact mechanism are largely independent: a phishing email can lead to either direct ICS-protocol abuse (Industroyer) or to ransomware-induced OT shutdown (Colonial Pipeline) depending on the attacker's objective and OT segmentation. \emph{Second}, the highest-consequence incidents (Stuxnet, Triton, INCONTROLLER) all required deep, vendor-specific knowledge of the target OT stack. This raises the bar on both attacker tradecraft and defender visibility, and motivates the AI/ML, Zero Trust, and digital-twin defensive technologies surveyed in Section~\ref{sec:tools_gaps}.

\section{Tools and Gaps in OT/IT Cybersecurity}
\label{sec:tools_gaps}

This section reviews the tooling landscape for defending OT and IT--OT environments and surfaces the gaps that remain. We organize the discussion around five tool categories---signature- and rule-based detection, AI/ML-driven detection, architectural defenses (Zero Trust, segmentation, blockchain logging, digital twins), incident-response and threat-intelligence tooling, and standards-aligned governance---and conclude with an explicit gap analysis tied back to the research questions in Section~\ref{sec:introduction}.

\subsection{Signature- and Rule-Based Detection}
Signature- and rule-based intrusion detection is the longest-deployed defensive layer in both IT and OT. Open-source platforms such as Snort, Suricata, and Zeek match network traffic against curated signatures of known attacks and protocol anomalies; host-based extensions (OSSEC, Wazuh) and malware-artefact scanners (YARA) extend the same idea to engineering workstations and historians. These tools detect known threats with high precision and have an established operational footprint, but they share two limitations relevant to OT: they require an attack signature to exist (a poor fit for novel ICS-specific malware such as Triton or INCONTROLLER), and their rule sets are predominantly IT-oriented---few rules natively understand industrial protocol semantics such as Modbus function codes or DNP3 object types. ICS-aware extensions exist (e.g., Modbus and DNP3 parsers in Suricata, ICS-protocol-aware rule packs maintained by ICS-CERT and security vendors), but coverage remains uneven.

\subsection{AI/ML-Driven Anomaly Detection}
Machine-learning approaches address the principal weakness of signature-based detection: the inability to recognize previously unseen attacks. Three families dominate the literature.

\textbf{Supervised classifiers} trained on labelled datasets such as KDD'99, NSL-KDD, UNSW-NB15 \cite{garcia2009anomaly, moustafa2016evaluation}, and ICS-specific datasets (SWaT \cite{goh2016swat}, WADI \cite{ahmed2017wadi}, HAI \cite{shin2020hai}) achieve high in-distribution accuracy but degrade on traffic patterns absent from training data.

\textbf{Unsupervised and semi-supervised anomaly detectors}---autoencoders, isolation forests, one-class SVMs, LSTM-based time-series predictors---learn baselines of normal device and protocol behavior and flag deviations \cite{mirsky2018kitsune}. They are well suited to OT because device behavior is comparatively stationary, but tuning false-positive rates remains a practical challenge in environments with frequent, legitimate process changes \cite{vinayakumar2019deep}.

\textbf{Hybrid AI plus signature pipelines} combine signature-based detection of known threats with ML-based anomaly detection for unknown threats. Reported evaluations consistently show that hybrids achieve higher detection rates and lower false-positive rates than either approach alone \cite{DBLP:journals/corr/abs-1807-07282}.

Rather than display point-precision accuracy, false-positive, and latency numbers---which would imply head-to-head benchmark results that the heterogeneous source studies do not support---we summarize the qualitative comparison across these method families in Table~\ref{tab:performance_comparison}. The cited references provide individual benchmarks on specific datasets (KDD'99, NSL-KDD, UNSW-NB15, SWaT, WADI) and tuning configurations; readers seeking specific numeric results should consult those references directly for the dataset and threat-model context they require.

\begin{table}[htbp]
\centering
\footnotesize
\caption{Qualitative comparison of intrusion-detection method families relevant to OT, drawn from the cited reference studies. Bands (L, M, H, VH) indicate cross-study consensus, not benchmark measurements; entries should be read as relative orderings rather than absolute performance numbers. Detailed numeric results vary substantially with dataset and tuning and are reported in the cited primary studies.}
\label{tab:performance_comparison}
\begin{tabular}{|p{3.0cm}|c|c|c|p{4.8cm}|}
\hline
\textbf{Method} & \makecell{\textbf{Detection}\\\textbf{accuracy}} & \makecell{\textbf{False-positive}\\\textbf{rate}} & \makecell{\textbf{Inference}\\\textbf{latency}} & \textbf{Adaptability and applicability} \\
\hline
Rule-based IDS (Snort, Suricata) & M & H & H & Low adaptability; misses zero-day and novel ICS-specific threats \cite{garcia2009anomaly} \\
\hline
Signature-based IDS (YARA, OSSEC) & M--H & M--H & M--H & Medium; bounded by signature database currency and ICS-protocol coverage \\
\hline
AI-based anomaly detection (ML/DL) & H & L--M & M & High; learns evolving patterns; sensitive to drift; explainability concerns for safety-critical alerts \cite{vinayakumar2019deep, mirsky2018kitsune} \\
\hline
Hybrid AI + Signature & H--VH & L & L--M & Very high; combines pattern learning and known-threat coverage; reported empirically superior to either alone in benchmarked studies \cite{DBLP:journals/corr/abs-1807-07282} \\
\hline
\end{tabular}
\end{table}

\noindent (Bands: L = low, M = medium, H = high, VH = very high. For accuracy: higher is better. For false-positive rate and latency: the band indicates magnitude, so L is best.)

\textbf{Operational caveats for OT.} (1)~ML models drift as devices, processes, or production schedules change; periodic retraining is required. (2)~Labelled OT attack datasets are scarce; SWaT, WADI, HAI, and Edge-IIoTset are commonly used but cover narrow scenarios. (3)~Real-time inference must respect deterministic cycle-time budgets, particularly on safety-relevant flows. (4)~Explainability matters: operators are unwilling to trust black-box alerts that interrupt physical processes.

\subsection{Architectural Defenses}

\textbf{Zero Trust Architecture (ZTA).} ZTA replaces implicit network-based trust with continuous, identity- and context-based verification, formalized in NIST Special Publication~800-207 \cite{nist80207, syed2022zero}. In OT, ZTA is most often deployed at the IT--OT boundary and within engineering networks: identity-aware proxies front jump servers, micro-segmentation restricts east--west movement between control zones, and adaptive authentication tightens controls when context (geolocation, device posture, time-of-day) is anomalous \cite{russoindustrial}. Industrial-automation vendors increasingly incorporate ZTA primitives such as role-based access control and continuous user verification into engineering workflows.

\textbf{Network segmentation and industrial DMZs.} Segmentation aligned with the Purdue Enterprise Reference Architecture \cite{williams1992purdue} remains the dominant defensive model. Industrial DMZs hosting historian replicas, jump servers, and patch-distribution servers constrain bidirectional traffic. Unidirectional security gateways (data diodes) provide hardware-enforced one-way egress from OT to IT for telemetry and historian replication \cite{heo2016design}.

\textbf{Blockchain-based event logging.} Tamper-evident logging of firmware updates, configuration changes, and operator commands using append-only ledgers addresses log-integrity concerns in environments where compromised hosts can rewrite their own audit trails \cite{javed2023blockchain, baggio2020blockchain}. Adoption remains limited and is concentrated in regulated power and pharmaceutical settings.

\textbf{Digital twins.} Real-time virtual replicas of physical OT systems support attack simulation, what-if analysis, and recovery rehearsal without disrupting production \cite{wang2023survey}. The Port of Rotterdam has implemented a port-wide digital-twin programme covering navigation, hydrography, and operations \cite{portrotterdam_dt, wang2021multi} which has been cited in the literature as a reference platform for cyber-incident rehearsal use cases. Sector-specific co-simulation testbeds for ICS cybersecurity research extend the same idea to lower-cost research settings: PULSE provides a configuration-driven co-simulation framework for microgrid industrial cyber-physical systems \cite{vardhan2026pulse}, NUCLEUS provides an open-source nuclear-power-plant simulator targeted at ICS-cybersecurity research \cite{vardhannucleus}, and complementary work has demonstrated real-time CIA-impact quantification on power-grid ICS using autonomous red-versus-blue agent loops \cite{vardhanquantifying}. Twins and co-simulation platforms together support synthetic data generation for ML training, partially mitigating the dataset-scarcity problem.

\textbf{Lightweight cryptography and secure communication.} Resource-constrained field devices motivate lightweight cipher suites and TLS profiles that fit within PLC and RTU compute budgets \cite{abir2024industry}. TLS~1.3 has progressively replaced legacy SSL/TLS in modernized substations and DCS deployments \cite{restuccia2020low}, and OPC~UA brings authenticated, encrypted communication into the application layer.

\subsection{Incident Response and Threat Intelligence}
Mature OT-aware Security Operations Centers integrate IT-side Security Information and Event Management (SIEM) platforms with dedicated OT-traffic monitoring and ICS-aware threat-intelligence feeds. Security Orchestration, Automation, and Response (SOAR) tooling automates containment actions---credential revocation, host isolation, traffic redirection---to reduce mean time to contain. Forensics tooling tailored for OT, including industrial-protocol dissectors and vendor-specific PLC-program diff tools, supports post-incident reconstruction. Deception assets---decoy PLCs, simulated Modbus/TCP endpoints, fake HMIs---increasingly complement detection by providing high-fidelity early warning of lateral movement.

Threat intelligence specific to OT adversary clusters (e.g., Sandworm, Xenotime, Electrum, Chernovite) is curated by specialist OT-security vendors (notably Dragos, Mandiant, Claroty), by Computer Emergency Response Teams (CERT-UA, US CERT/CISA), and by national security agencies (NSA, NCSC). MITRE ATT\&CK for ICS \cite{ICSATTACK} provides a standardized vocabulary for adversary tactics and techniques against ICS, complementing the enterprise ATT\&CK matrix.

\subsection{Architectural Resilience: A Six-Domain View}
\begin{figure}[htbp]
    \centering
    \includegraphics[width=0.85\linewidth]{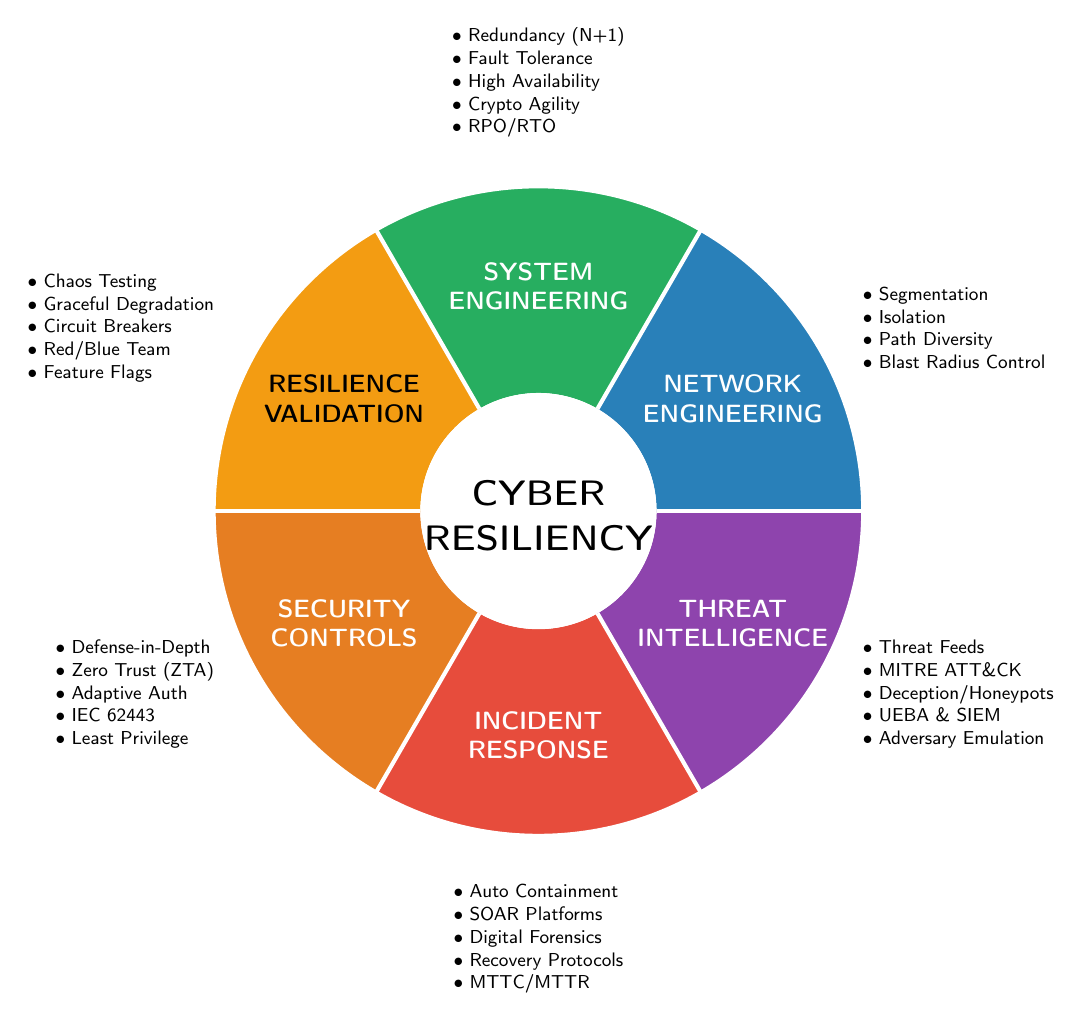}
    \caption{Six-domain decomposition of OT cyber resilience used in this survey, with representative practices for each domain. The decomposition integrates techniques drawn from NIST SP~800-160 Vol.~2 and MITRE's Cyber Resiliency Engineering Framework; the six-domain partition itself is the authors' synthesis.}
    \label{fig:cyber_resiliency}
\end{figure}

Beyond individual tools, several practitioner frameworks organize OT defenses across architectural domains. NIST SP~800-160 Vol.~2 \cite{nistsp800160v2} structures cyber-resilience around the four functions \emph{Anticipate, Withstand, Recover, Adapt} and a set of techniques (segmentation, redundancy, diversity, deception, adaptive response) that map onto network, system, and operational layers. MITRE's Cyber Resiliency Engineering Framework Navigator \cite{mitre_cref} catalogues 14 techniques and 50+ approaches against the same functions. Building on these foundations, Figure~\ref{fig:cyber_resiliency} and Table~\ref{tab:resilience_layers} present a layered six-domain decomposition---network engineering, system engineering, resilience validation, security controls, incident response, and threat intelligence---that we use as an organizing pattern for the remainder of the section. The decomposition is our synthesis rather than a citation from either framework; the intent is to ensure that when one layer fails, others continue to constrain attacker progress and preserve essential function.

\begin{table}[htbp]
\centering
\footnotesize
\caption{Architectural domains for OT cyber resilience and representative practices (companion to Figure~\ref{fig:cyber_resiliency}).}
\label{tab:resilience_layers}
\renewcommand{\arraystretch}{1.2}
\begin{tabular}{|p{3.2cm}|p{11.5cm}|}
\hline
\textbf{Domain} & \textbf{Representative practices} \\ \hline
Network Engineering & Segmentation (VLANs, micro-segments, zones aligned to ISA/IEC 62443); isolation via firewalls, SDN policies, or unidirectional gateways; path diversity for survivability. \\ \hline
System Engineering & Redundancy ($N+1$, $N+2$) for PLCs, SCADA servers, and links; quorum-based HA for control planes; warm/cold standby for stateful tiers; cryptographic agility (multiple CAs, staggered rotation); short RPO/RTO via immutable backups, PITR, IaC rebuilds. \\ \hline
Resilience Validation & Chaos and failure testing (game days, fault injection, packet loss/jitter, CA/KMS outages); graceful degradation patterns (feature flags, circuit breakers, read-only modes, queue buffering, rate limiting). \\ \hline
Security Controls & Defense-in-depth aligned with IEC~62443 zones and conduits; least-privilege and continuous-verification access via ZTA; adaptive authentication based on contextual risk. \\ \hline
Incident Response & SOAR-driven automated containment; integrated forensics (logs, memory captures, endpoint telemetry); validated, IaC-driven recovery from clean baselines. \\ \hline
Threat Intelligence \& Adaptation & MITRE ATT\&CK for ICS-aligned analytics; OT adversary profiles; deception assets (decoy PLCs, fake HMIs); UEBA tuned to OT telemetry (setpoint deltas, command-rate deviations). \\ \hline
\end{tabular}
\end{table}

For organizations seeking quantitative tracking of these domains, common operational metrics include Mean Time to Detect (MTTD), Mean Time to Contain (MTTC), and Mean Time to Recover (MTTR), all of which are well established in the SOC and SRE practitioner literature. We additionally define two working metrics for this paper: \emph{Residual Functional Capacity} (RFC, the percentage of mission functions preserved during disruption) and \emph{Blast Radius Coefficient} (BRC, the proportion of dependent nodes affected by a localized fault); these are inspired by concepts in the resilience and reliability-engineering literature on graceful degradation and fault propagation rather than NIST or MITRE-defined metrics. Composite indices that aggregate these into single scores have been proposed in the literature \cite{nistsp800160v2, mitre_cref} but lack empirical validation across diverse OT settings; we treat them as research artifacts rather than deployable tools.

\subsection{Standards and Regulatory Landscape (Pointer)}
Governance of OT cybersecurity rests on a set of overlapping U.S., European, and international frameworks (NIST CSF~2.0, NIST~SP~800-82~Rev.~3, IEC~62443, EU NIS2, DORA, the EU Cyber Resilience Act, NERC CIP, and sector-specific regimes for healthcare, aviation, and finance). Section~\ref{sec:reg_frameworks}, after the historical and commercial-impact analysis of Section~\ref{sec:history_impact}, treats these in detail. For the purposes of this section, the relevant observation is that all major frameworks now expect---explicitly or by implication---the tooling categories surveyed above (segmentation, identity-centric access, ML-augmented monitoring, supply-chain attestation). The principal practitioner question has shifted from \emph{whether} to deploy these tools to \emph{which} cross-framework cross-walk to use as the audit anchor.

\subsection{Gaps and Open Problems}
\label{subsec:gaps}
Despite the breadth of available tooling, the surveyed literature and incident record reveal seven gaps where research and engineering investment is most needed.

\begin{enumerate}
    \item \textbf{OT-aware patch management.} Tooling for safe, validated, in-context patching of PLCs and DCS controllers---including pre-deployment regression testing against control-loop models---is immature. Long revalidation cycles leave known vulnerabilities open for months to years.
    \item \textbf{Labelled OT attack datasets.} The community largely relies on a small number of testbed datasets (SWaT, WADI, HAI, Edge-IIoTset). Coverage of vendor-specific attack patterns, multi-stage campaigns, and physical-process effects is thin. Sector-specific co-simulation platforms---PULSE for microgrids \cite{vardhan2026pulse}, NUCLEUS for nuclear power plants \cite{vardhannucleus}, and CIA-impact frameworks driven by autonomous red--blue agents \cite{vardhanquantifying}---are an emerging complement that can generate labelled physical-process attack data at low cost and across vendor stacks not represented in existing testbeds.
    \item \textbf{ML drift and explainability for safety-critical alerts.} ML models retrained on shifting baselines must remain trustworthy enough that operators will act on alerts that interrupt physical processes. Methods for monitoring concept drift and producing operator-readable explanations are an active research area.
    \item \textbf{Authentication retrofit for legacy protocols.} Modbus, EtherNet/IP, and unsecured DNP3 deployments dominate the installed base. In-line authenticators, proxy gateways, and protocol-aware firewalls that retrofit authentication without controller replacement are needed.
    \item \textbf{IoMT and healthcare OT segmentation.} Hospitals routinely run flat or weakly segmented networks across clinical IT, IoMT, and facility OT. AI-driven traffic clustering is starting to inform Zero Trust zones \cite{ibm2025xforce, claroty2024healthcare, nist1800-8}, but engineered, vendor-supportable IoMT micro-segmentation is far from standard.
    \item \textbf{Consistent resilience metrics.} Multiple frameworks (NIST SP~800-160 Vol.~2, DoD CREM, MITRE Cyber Resiliency Engineering, IEC~62443 SL, NERC CIP) define overlapping but inconsistent resilience measures. A common, instrumentable, cross-framework set of resilience metrics tied to operational telemetry would advance both research and practice.
    \item \textbf{Supply-chain attestation for firmware and software.} SBOM adoption, signed firmware, and runtime attestation of PLC and RTU firmware are still nascent. Trusted-execution-environment-backed attestation in industrial controllers is a clear research direction.
\end{enumerate}

These gaps inform the open-problems discussion in Section~\ref{sec:conclusion} and motivate the historical impact analysis that follows in Section~\ref{sec:history_impact}: where defensive tooling falls short, attackers have demonstrated their willingness to exploit the gap, often at substantial commercial cost.

\section{Attack History and Commercial Effects}
\label{sec:history_impact}

This section compiles the historical record of high-impact OT cyber incidents from 2010 through 2025 and quantifies their commercial effects. The case studies illustrate, in concrete terms, how the attack vectors taxonomized in Section~\ref{sec:ga_sbo} have manifested in practice, and the cost analysis grounds the gap discussion of Section~\ref{sec:tools_gaps} in financial reality. Where numerical claims are reported, we attribute them to their issuing organization and report year; estimates derived from worst-case modelling or industry surveys are flagged accordingly.

\subsection{Timeline of Notable Incidents}
\begin{figure}[h]
    \centering
    \includegraphics[width=0.9\linewidth]{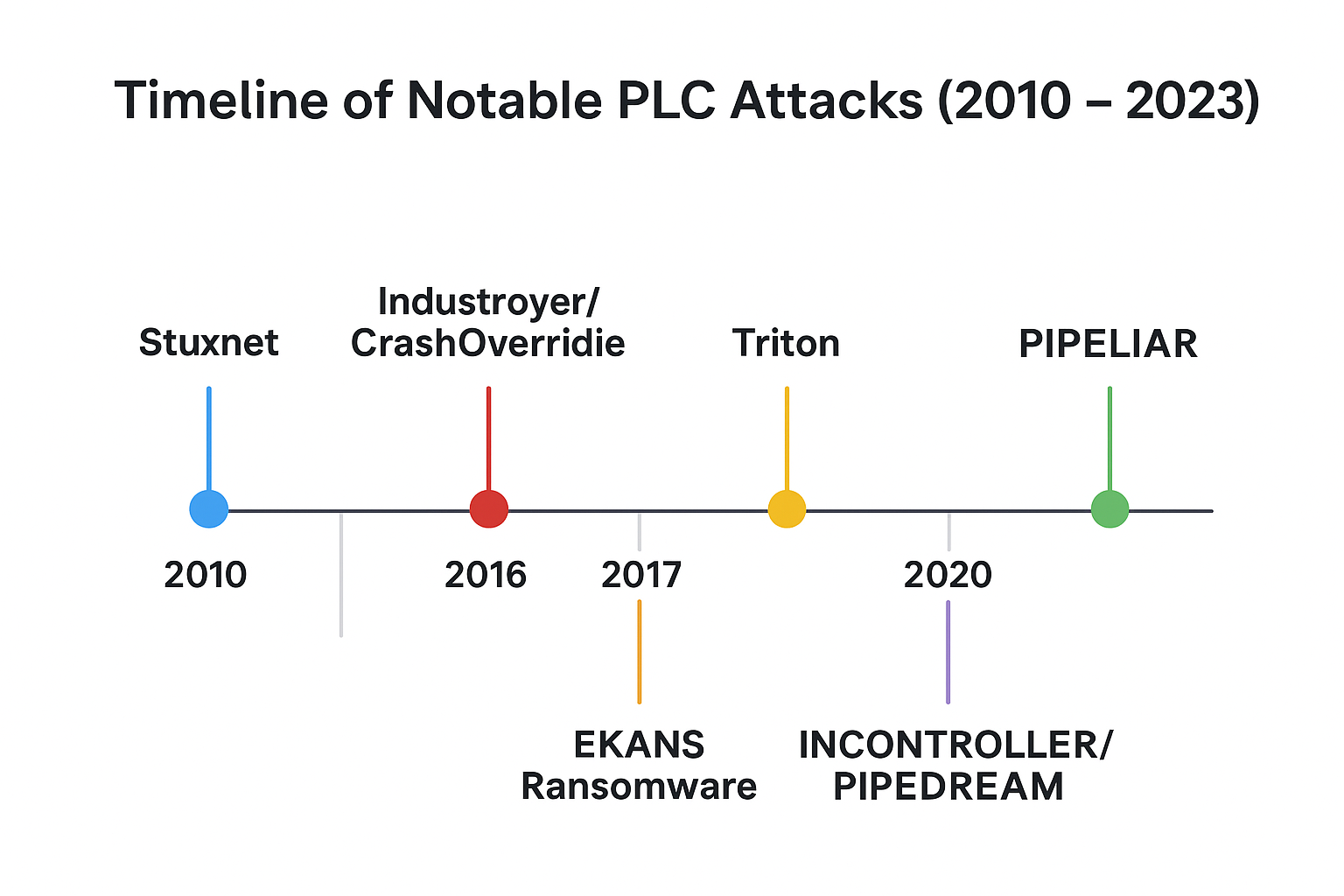}
    \caption{Timeline of notable PLC- and ICS-targeted attacks, 2010--2023. A comprehensive 69-incident catalogue spanning 2010--2025 is provided in Table~\ref{tab:full_timeline}.}
    \label{fig:plc_timeline}
\end{figure}

We organize the case-study evidence into three disjoint eras. \emph{Era~I (2010--2016)} covers the foundational ICS-targeted operations: bespoke nation-state malware (Stuxnet, Duqu, Flame), the wiper-as-strategic-disruption pattern (Shamoon), and the first publicly attributed cyber-induced civilian power outage (Ukraine 2015 / Industroyer 2016). \emph{Era~II (2017--2021)} covers the maturation of cascading wormable IT events with collateral OT impact (WannaCry, NotPetya, TSMC), the first SIS-targeted attack (Triton), and the rise of OT-aware ransomware (Norsk Hydro, EKANS, JBS, HSE Ireland, Colonial Pipeline). \emph{Era~III (2022--2025)} covers wartime ICS operations (AcidRain, Industroyer2, FrostyGoop, Fuxnet), modular ICS toolkits in the wild (INCONTROLLER), low-bar opportunistic PLC abuse (CyberAv3ngers / Aliquippa), and a new wave of large-financial-impact ransomware (Toyota/Kojima, CommonSpirit, Change Healthcare, Ascension, CDK Global, JLR). After the case studies we present a comprehensive cross-validated incident table (Table~\ref{tab:full_timeline}) and discuss how the direct-ICS versus ransomware-on-OT-operators distinction reshapes interpretation of the historical record.

\subsection{Era I: Foundational Incidents (2010--2016)}

\textbf{Stuxnet (2010) --- Natanz, Iran.}
Stuxnet was the first widely analyzed cyber-physical attack \cite{Langner, kushner2013real, farwell2011stuxnet, CCDCOE}. Targeting Siemens S7-300 and S7-400 PLCs (specifically the S7-315 and S7-417 models) that controlled uranium-enrichment centrifuges, the worm propagated via removable USB drives, exploited four Microsoft Windows zero-day vulnerabilities to traverse engineering networks, and reprogrammed PLC logic to drive centrifuges through resonance speeds while reporting nominal values to the operator HMI. The Institute for Science and International Security (ISIS) estimated in December 2010 that the operation may have damaged up to approximately 1{,}000 IR-1 centrifuges (about 10\% of the operating stock) at the Natanz Fuel Enrichment Plant between November 2009 and late January 2010 \cite{isis2010stuxnet}. Stuxnet established the precedent for nation-state Advanced Persistent Threats targeting critical infrastructure \cite{stojanovic2020apt}.

\textbf{Shamoon (2012) --- Saudi Aramco and RasGas.}
On 15 August 2012, the Shamoon wiper destroyed approximately 30{,}000--35{,}000 workstations at Saudi Aramco, with a parallel incident at RasGas in Qatar two weeks later \cite{shamoon2012}. While Shamoon targeted IT systems rather than ICS directly, the operational fallout---including disruption to drilling, refining, and shipping coordination---demonstrated the cascading effect of large-scale IT wiper events on energy operations. Shamoon variants reappeared in 2016 (Shamoon~2) and 2018 (Shamoon~3, against the Italian oil-services company Saipem). Attribution has consistently pointed to Iran-aligned actors. Shamoon established the wiper-as-strategic-disruption pattern that NotPetya, AcidRain, and CaddyWiper would later inherit.

\textbf{Havex / Dragonfly (2014) --- Energy and Pharmaceutical Espionage.}
Havex, deployed by the actor cluster Symantec named Dragonfly (later Energetic Bear), was distributed via trojanized installer downloads on the websites of three industrial-control-system vendors \cite{havex2014}. Once installed on engineering workstations, Havex enumerated OPC servers and scanned local ICS networks for exploitable devices. While no destructive payload was deployed, Havex demonstrated that adversaries could compromise the supply chain of trusted ICS software to fingerprint hundreds of energy and pharmaceutical operators across the U.S. and Europe---an espionage-and-pre-positioning operation that foreshadowed the supply-chain dimension of INCONTROLLER eight years later.

\textbf{German Steel Mill (December 2014) --- Blast Furnace Damage.}
The German Federal Office for Information Security (BSI) reported in its 2014 \emph{Lagebericht} that a (still unnamed) German steel mill suffered a cyberattack that prevented controlled shutdown of a blast furnace, resulting in ``massive'' physical damage to the furnace \cite{germansteel2014}. Initial access was via spear-phishing into the office IT network, with subsequent lateral movement to plant-floor systems. The German steel mill incident is the second publicly documented case (after Stuxnet) in which a cyberattack produced confirmed destructive physical damage to industrial equipment, and it remains the canonical example outside the nation-state/critical-infrastructure context.

\textbf{BlackEnergy / Ukraine 2015 --- Power Distribution.}
On 23 December 2015, three Ukrainian regional electricity-distribution companies (oblenergos) lost dispatcher control after attackers entered the IT network via spear-phishing, harvested credentials over months, and ultimately operated SCADA HMIs directly to open breakers \cite{ukraine2015cisa, ukraine2015sans}. Approximately 225{,}000 customers lost power for one to six hours; attackers also deployed KillDisk to hinder recovery and a TDoS attack on the dispatch call centre. This was the first publicly attributed cyberattack to cause widespread civilian power loss.

\textbf{PLC-Blaster (Black Hat Asia 2016) --- Worm Targeting Siemens PLCs.}
PLC-Blaster, presented by Spenneberg, Br\"uggemann, and Schwartke at Black Hat Asia 2016 \cite{spenneberg2016plcblaster}, is a proof-of-concept worm that demonstrated autonomous PLC-to-PLC propagation across an industrial network. After initial infection of a Siemens S7-1200 PLC, the worm scanned the OT network for additional vulnerable controllers and replicated without human intervention. While PLC-Blaster did not produce a major commercial incident, it formalized the threat model of self-propagating ICS malware that bypasses engineering workstations entirely.

\textbf{Industroyer / CrashOverride (2016) --- Kyiv, Ukraine.}
A year after the 2015 incident, Industroyer struck Ukrenergo's Pivnichna transmission substation in northern Kyiv on 17 December 2016, with attribution to the Sandworm group \cite{kozak2023industroyer, ESET, dragos, sans}. Industroyer differed from BlackEnergy in that it natively spoke ICS protocols (IEC~60870-5-101/104, IEC~61850, OPC~DA), enabling direct manipulation of circuit breakers without HMI intermediation. The malware also included a data-wiper module to hinder recovery. The attack caused approximately one hour of outage affecting roughly one-fifth of Kyiv. Industroyer demonstrated that adversaries had moved from operating ICS via stolen HMIs to embedding ICS protocol logic directly in malware.

\subsection{Era II: Wormable Cascades, SIS-Targeted Malware, and OT-Aware Ransomware (2017--2021)}

\textbf{WannaCry (May 2017) --- Wormable IT Cascade into OT.}
WannaCry combined a leaked NSA exploit (EternalBlue, targeting Microsoft SMBv1) with self-propagating ransomware to infect more than 200{,}000 machines in over 150 countries within days \cite{wannacry2017nao}. While WannaCry was not designed to target OT, its self-propagation through unpatched Windows endpoints reached the OT-adjacent floor in multiple sectors: Renault and Nissan halted production lines; Honda paused its Sayama plant; Boeing reported affected systems; the UK National Audit Office reported that NHS England confirmed 6{,}912 cancellations of appointments and operations during incident management (12--18 May 2017) and estimated more than 19{,}000 cancellations in total once unrecorded primary-care impact was extrapolated \cite{wannacry2017nao}. Attribution by the U.S. and UK governments points to the DPRK-linked Lazarus group. WannaCry is the canonical example of an IT-side ransomware worm producing severe OT-adjacent operational consequences purely through collateral propagation.

\textbf{NotPetya (June 2017) --- The \$10 Billion Wiper.}
NotPetya was distributed via a trojanized update to the Ukrainian accounting software M.E.Doc and rapidly propagated through corporate networks worldwide using EternalBlue and credential-theft techniques \cite{notpetya2018wh}. Unlike WannaCry, NotPetya's encryption was unrecoverable by design---a wiper masquerading as ransomware. The financial damage is the largest publicly documented for any cyber event. Maersk, the world's largest container shipping line, guided to approximately US\$200--300 million in Q3-2017 revenue impact in its August 2017 trading update \cite{maersk2017q2update}; 17 of the 76 port and terminal facilities operated by Maersk's APM Terminals subsidiary reverted to manual operation \cite{maritimeexec_maersk2017}; Merck disclosed US\$870 million; FedEx/TNT US\$400 million; Saint-Gobain US\$384 million; Mondelez US\$150--188 million. Total global damages are estimated at approximately US\$10 billion. The U.S., UK, and EU governments attribute NotPetya to Sandworm (Russia GRU). NotPetya remains the most-cited evidence that IT-side cascades can produce strategically significant OT outcomes without ICS-specific tooling.

\textbf{Triton / Trisis / HatMan (2017) --- Petrochemical Plant, Saudi Arabia.}
Triton targeted Schneider Electric's Triconex Safety Instrumented System (SIS) at a petrochemical facility in Saudi Arabia \cite{mekdad2021threat, trisis_drago}. The attackers compromised an engineering workstation connected to the SIS network and attempted to reprogram the SIS controllers, either to disable safety functions or to induce unsafe states. The attack was detected when a Triconex unit faulted and the plant tripped to a safe state. Triton was the first publicly known cyberattack to specifically target a SIS---the last automated line of defense against catastrophic process failures---and remains the highest-consequence near-miss in OT security history.

\textbf{TSMC (August 2018) --- WannaCry Variant Crashes Semiconductor Fab.}
On 3 August 2018, Taiwan Semiconductor Manufacturing Company experienced a self-propagating WannaCry variant that crashed approximately 10{,}000 fab-automation hosts running unpatched Windows~7 \cite{tsmc2018}. The infection vector was a tool brought onto the fab LAN without prior antivirus scanning. TSMC initially estimated approximately US\$170 million (then 3\% of Q3 2018 revenue) in revenue impact \cite{tsmc2018}; the company subsequently booked NT\$2.6 billion ($\approx$~US\$85 million at 2018 exchange rates) of cyber-incident-related losses in its Q3/Q4 2018 financial statements \cite{tsmc2018booking}. The TSMC incident illustrates how unpatched legacy operating systems on fab-automation equipment---the OT analog of long-lifecycle PLCs---become single points of failure for high-throughput, time-sensitive industrial operations.

\textbf{Norsk Hydro / LockerGoga (March 2019) --- Manual Operation of Aluminum Smelters.}
On 18--19 March 2019, the LockerGoga ransomware encrypted approximately 22{,}000 computers across 170 sites in roughly 40 countries operated by the Norwegian aluminum producer Norsk Hydro \cite{hydro_cyberpage, norskhydro2019}. The company's response became a sector-wide reference case: Norsk Hydro publicly refused to pay the ransom, switched aluminum smelters and rolling mills to manual operation guided by retired employees, and disclosed daily progress through public press conferences. Hydro disclosed total estimated losses of NOK 650--750 million ($\approx$ US\$67--84 million) in its 2019 annual report \cite{hydro_ar2019}. The Norsk Hydro incident demonstrated that for some industrial processes manual fallback is feasible at substantial cost, and established the transparent disclosure pattern that several later high-profile victims (Maersk, Bridgestone) would emulate.

\textbf{EKANS / Snake Ransomware (Dec 2019; reported 2020) --- Manufacturing.}
EKANS (also known as Snake), first observed in late December 2019 and publicly analyzed in early 2020, is ransomware that incorporated explicit awareness of OT environments. Initial access was typically via phishing or exposed RDP. Once on the network, EKANS used legitimate administrative tools (PowerShell, PsExec, WMI) for lateral movement, then terminated a hard-coded list of ICS-related processes (data historians, OPC servers, vendor-specific applications) before encrypting files. Honda reported a June 2020 cyber incident that interrupted manufacturing at multiple plants; published analyses noted the techniques resembled EKANS. While later analyses (Dragos, 2020) clarified that EKANS' OT-process kill list reflected awareness of OT environments rather than active control of physical processes, the malware nonetheless marked the point at which financially motivated ransomware operators began deliberately tailoring payloads against OT-adjacent systems.

\textbf{Colonial Pipeline (May 2021) --- Fuel Distribution, U.S. East Coast.}
The Colonial Pipeline ransomware incident in May 2021 illustrates how an IT-side compromise can produce massive OT-side commercial impact even without direct OT manipulation. Attackers (the DarkSide ransomware affiliate) entered through a legacy VPN account that lacked multi-factor authentication and encrypted IT systems. Colonial proactively shut down the pipeline---which transports approximately 45\% of U.S. East Coast fuel---to contain the incident; the pipeline was offline for approximately five days (7--12 May 2021) with all systems and operations reported normal by 15 May, producing regional fuel shortages and an emergency federal response \cite{guardian2021colonial, cisa2023colonial}. Colonial paid a ransom of approximately \$4.4 million (subsequently partially recovered by the U.S.\ Department of Justice); widely reported industry estimates placed daily losses during the shutdown above \$25 million.

\textbf{JBS Foods (May 2021) --- Global Meat Supply.}
On 30 May 2021, the REvil ransomware operators encrypted JBS Foods' IT systems across the U.S., Canada, and Australia \cite{jbs2021}. JBS halted operations at 13 plants representing approximately one-fifth of U.S. beef-processing capacity and a significant share of Australian and Canadian protein supply. The company paid an US\$11 million ransom in Bitcoin to limit further disruption. Coupled with Colonial Pipeline two weeks earlier, JBS contributed to a U.S. policy shift treating ransomware as a national-security concern and triggered the May 2021 Executive Order on Improving the Nation's Cybersecurity.

\textbf{HSE Ireland / Conti (May 2021) --- Healthcare-Wide Ransomware.}
On 14 May 2021 the Conti ransomware operators encrypted systems across Ireland's Health Service Executive following an eight-week dwell after a March 2021 phishing email; hospitals reverted to paper for weeks, and the official PwC post-incident review documented continuous recovery activity through late 2021 \cite{hsepostreview}. The HSE's own 2024 estimate places direct incident-response cost at approximately €102 million; the Irish Comptroller and Auditor General projects an additional approximately €657 million over a seven-year horizon for the resulting cybersecurity-improvement programme \cite{hse2024cag, rte2024hse}.

\subsection{Era III: Wartime ICS Operations, Modular Toolkits, and Cascading Ransomware (2022--2025)}

\textbf{Viasat KA-SAT and AcidRain (24 February 2022) --- Satellite Modem Wiper.}
Coinciding with Russia's invasion of Ukraine on 24 February 2022, attackers used VPN credential abuse against the management plane of the Viasat KA-SAT satellite-broadband network and deployed AcidRain, a wiper that destroyed modem firmware on tens of thousands of customer terminals across Europe \cite{acidrain2022}. The most cited collateral effect was the loss of remote monitoring and control of approximately 5{,}800 Enercon wind turbines in Central Europe (with a combined nameplate output of roughly 11\,GW, predominantly in Germany), per Enercon's 1 March 2022 statement \cite{enercon2022}---note that the turbines continued to spin and produce power, but operators could no longer adjust them remotely. The U.S., U.K., and E.U. attributed the attack to Russia GRU. AcidRain established a wartime precedent for cyber operations against satellite-coupled OT and remains the highest-profile wartime cyber-physical operation against civilian infrastructure during active armed conflict to date.

\textbf{Toyota / Kojima Industries (February 2022) --- Tier-1 Supplier Just-in-Time Failure.}
On 26 February 2022, Kojima Industries---a tier-1 Toyota supplier of plastic resin parts---detected unauthorized access to its systems; on 1 March 2022 Toyota suspended operations at all 14 of its Japanese assembly plants (28 production lines) for one day, restarting from the first shift on 2 March \cite{toyotakojima2022, computerweekly_toyota2022}. Approximately 13{,}000 vehicles were lost in a single day, equivalent to roughly one day of Toyota's domestic Japanese production and approximately 5\% of monthly production capacity per contemporaneous industry-analyst estimates. The incident illustrates a class of OT impact that is essentially indirect: Toyota's own OT was unaffected, but the just-in-time inventory model coupled it tightly enough to its supplier's IT health that the supplier's compromise produced the largest single-day production loss in Toyota's modern history.

\textbf{Industroyer2 (April 2022) --- Foiled Wartime Attack on Ukraine's Grid.}
On 8 April 2022, an attempted second-generation Industroyer attack against a Ukrainian high-voltage substation was foiled jointly by CERT-UA and ESET \cite{industroyer2_eset}. Industroyer2 was tailored to a specific substation rather than the modular protocol-agnostic design of the 2016 original, and it was paired with the CaddyWiper destructive payload. Sandworm intrusion preparation reportedly began in February 2022. The incident is the first publicly documented case of a near-miss ICS attack during active armed conflict.

\textbf{INCONTROLLER / PIPEDREAM (April 2022) --- Vendor-Agnostic ICS Toolkit.}
Disclosed by CISA, the U.S. Department of Energy, NSA, and FBI on 13 April 2022 in Joint Cybersecurity Advisory AA22-103A \cite{incontroller2022}, INCONTROLLER (also tracked as PIPEDREAM by Dragos) is a modular attack toolkit capable of issuing commands across multiple PLC and engineering-tool families, including Schneider Electric Modicon, OMRON Sysmac NEX, OPC~UA servers, and CODESYS-based controllers. It was discovered before deployment to a target. Dragos labels the operating actor cluster ``Chernovite'' but stops short of state attribution; the joint U.S.\ advisory is similarly cautious. INCONTROLLER represented a step-change in ICS-malware engineering: a reusable, vendor-agnostic platform comparable in posture to enterprise IT post-exploitation frameworks.

\textbf{CommonSpirit Health (October 2022) --- 100 Facilities, 13 States.}
A ransomware incident at one of the largest U.S.\ hospital systems---which operates approximately 140 hospitals---disrupted clinical and facility operations at roughly 100 facilities across 13 states; the company's SEC 10-K disclosed restoration and lost-revenue costs in the order of US\$160 million \cite{commonspirit10k}.

\textbf{CyberAv3ngers / Aliquippa Water (November 2023) --- Internet-Exposed PLCs.}
On 25 November 2023, attackers identified by CISA as ``CyberAv3ngers,'' an Iran IRGC-affiliated group, defaced the HMI of a Unitronics Vision-series PLC at the Municipal Water Authority of Aliquippa, Pennsylvania \cite{cyberavengers2023}. The booster station regulates pressure for the Raccoon and Potter Townships, downstream of an MWAA distribution serving approximately 6{,}600 customer accounts; staff placed the affected zone on manual fallback without water-quality impact. CISA's joint advisory AA23-335A (1 December 2023) documented compromises of internet-exposed Unitronics Vision-series PLCs across multiple U.S.\ critical-infrastructure sectors, with the water-and-wastewater sector most heavily represented. The Aliquippa incident is significant for its low technical bar: the attackers exploited internet-exposed Unitronics PLCs running default credentials, a finding that contradicts the widespread assumption that ICS attacks require sophisticated tradecraft.

\textbf{FrostyGoop / Lviv (January 2024) --- Modbus-TCP Native ICS Malware.}
Disclosed by Dragos in July 2024, FrostyGoop is the first publicly documented ICS-targeting malware to communicate natively over Modbus TCP \cite{frostygoop2024}. In January 2024, attackers used FrostyGoop to falsify temperature readings reported by district-heating controllers in Lviv, Ukraine, leaving approximately 600 apartment buildings without heat for two days in sub-zero conditions. Initial access was traced to an internet-exposed MikroTik router compromised in April 2023. FrostyGoop's significance is twofold: it confirms that adversaries are continuing to invest in protocol-native ICS tooling beyond the IEC-104 and Triconex precedents, and it produced direct civilian-welfare harm in a wartime context.

\textbf{Change Healthcare (February 2024) --- US\$2.46 Billion Healthcare Cascade.}
On 12 February 2024, the ALPHV/BlackCat ransomware group encrypted Change Healthcare, the U.S. payment-and-claims clearing house owned by UnitedHealth Group, after stealing credentials for a Citrix portal that lacked multi-factor authentication \cite{changehc2024}. The resulting cascade disrupted pharmacy claims, prior authorizations, and provider payments across the U.S. for weeks, with knock-on effects on patient care, prescription dispensing, and provider cash flow. UnitedHealth Group disclosed approximately US\$22 million paid in ransom, an estimated 190 million individuals impacted as of January 2025 (revised upward to approximately 192.7 million in the July 2025 OCR notification), and total cost of approximately US\$2.46 billion through 2024 (per UnitedHealth Group's Q3 2024 earnings)---among the highest single-event costs in U.S.\ cyber history \cite{changehc2024, hhs_changehc_faq}. The incident motivated congressional hearings and has been cited in support of mandatory MFA requirements for healthcare ICT third parties.

\textbf{Fuxnet / Moscow (April 2024) --- Sensor-Gateway Wiper.}
Analyzed by Claroty Team82 in April 2024, Fuxnet is wiper malware deployed by the pro-Ukraine ``Blackjack'' group against Moscow's Moscollector industrial sensor network \cite{fuxnet2024}. Fuxnet destroyed NAND and UBI flash on approximately 500 sensor gateways (with the threat actor claiming up to 87{,}000) and included a Modbus and M-Bus protocol fuzzer to maximize downstream damage. Fuxnet is notable as a wartime ICS operation \emph{against} Russian infrastructure---a counterpart to AcidRain and Industroyer2.

\textbf{Ascension Health (May 2024) --- US\$1.8 Billion FY24 Operating Loss.}
On 8 May 2024, the Black Basta ransomware group encrypted systems across Ascension Health, one of the largest U.S.\ Catholic healthcare providers, after an employee downloaded a malicious file \cite{ascension2024}. Electronic Health Records, laboratory systems, and radiology workflows were taken offline at 142 hospitals across 19 states (and the District of Columbia); emergency-department diversions were widespread. Ascension's HHS OCR breach-portal filing was updated in December 2024 to approximately 5.6 million records affected (5{,}599{,}699 individuals), and Ascension reported an FY24 operating loss of approximately US\$1.8 billion, to which the cyber-incident response was a significant contributor.

\textbf{Synnovis / Qilin (June 2024) --- NHS Pathology Lab Encrypted.}
On 3 June 2024, the Qilin ransomware operators encrypted the IT systems of Synnovis, a pathology service provider for several NHS trusts in London \cite{synnovis2024}. Within 13 days of the incident, NHS England reported 1{,}134 cancelled operations and 2{,}194 cancelled outpatient appointments across 7 hospitals; approximately 400~GB of patient data was leaked to Qilin's Telegram channel. NHS England subsequently identified contributory factors to one patient death. Synnovis remains a defining case study for the patient-safety implications of ransomware against shared clinical-IT third parties.

\textbf{CDK Global (June 2024) --- Auto-Dealer Operations.}
On 18 June 2024, the BlackSuit ransomware group encrypted CDK Global's dealer-management SaaS platform \cite{cdkglobal2024}. Approximately 15{,}000 U.S. and Canadian auto dealerships reverted to paper-based sales, service, and parts processes from 19 June through approximately 5 July 2024. Anderson Economic Group (AEG) estimated direct dealer losses at approximately US\$1.02 billion across the three-week disruption (lost earnings on new and used unit sales, parts and service, plus additional staffing/IT and floor-plan-interest costs) \cite{aeg2024cdk}; CDK reportedly paid a US\$25 million ransom. CDK is a sectoral analog of Toyota/Kojima: a critical OT-adjacent SaaS provider whose compromise multiplied across thousands of downstream operators.

\textbf{M\&S, Co-op, Harrods (April 2025) --- DragonForce / Scattered Spider.}
Beginning 19 April 2025, U.K.\ retailers Marks \& Spencer, Co-op, and Harrods were targeted by helpdesk-based social-engineering attacks (the Scattered Spider technique cluster, also tracked as UNC3944) using the DragonForce ransomware-as-a-service platform \cite{ms_coop_2025}. M\&S online ordering was offline for weeks and physical food-supply distribution was disrupted. The UK Cyber Monitoring Centre classified the combined M\&S--Co-op event as a Category~2 incident on its hurricane-style scale, with combined modelled financial impact of approximately £270--440 million across the two retailers \cite{cmc_msccoop2025}; M\&S separately warned investors of approximately £300 million in profit impact for the year. Four arrests were announced by the UK National Crime Agency in July 2025.

\textbf{Jaguar Land Rover (August--October 2025) --- Most Costly UK Cyberattack.}
On 31 August 2025, attackers (claiming the ``Scattered Lapsus\$ Hunters'' identity, an apparent collaboration of Lapsus\$, Scattered Spider, and ShinyHunters) compromised Jaguar Land Rover, halting all UK production lines for approximately five weeks. JLR lost approximately 5{,}000 vehicles per week with modelled losses to JLR's UK manufacturing operations of approximately £108 million per week (fixed costs plus lost profit); the UK Cyber Monitoring Centre's modelled estimate of £1.9 billion in total UK economy impact---affecting more than 5{,}000 UK organisations through supplier ripple effects---makes JLR the most costly cyberattack in UK history to date \cite{cmc_jlr2025}. JLR demonstrates the same just-in-time amplification effect as Toyota/Kojima but at significantly greater scale.

\textbf{Iran-Affiliated PLC Campaign (2025--2026) --- U.S. Water and Energy.}
Throughout 2025 and into 2026 an Iran-affiliated campaign continued against internet-exposed PLCs across U.S.\ water-and-wastewater utilities, energy facilities, and government installations, formalised in joint Cybersecurity Advisory AA26-097A issued 7 April 2026 by FBI, CISA, NSA, EPA, the U.S.\ Department of Energy, and U.S.\ Cyber Command \cite{iranplc2026}. AA26-097A specifically calls out exploitation of Rockwell Automation / Allen-Bradley programmable logic controllers (extending the prior 2023 CyberAv3ngers / Unitronics pattern to a different vendor stack) by IRGC-CEC--linked actors operating under the CyberAv3ngers persona, with disruption to victim HMI and SCADA systems and operational/financial impact across multiple sectors. Separately, in December 2023 a small Irish water utility serving an Irish village experienced approximately two days without water service following exploitation of an internet-exposed Unitronics device.

\textbf{Other Notable Incidents.}
\begin{itemize}
    \item \emph{Oldsmar Water Treatment (2021), disputed.} A change in sodium hydroxide setpoint was observed on the HMI of a Florida water utility's TeamViewer-connected workstation; an operator reverted the change before any treatment effect occurred. The event was initially reported as an external intrusion, but subsequent statements from city officials in 2023 raised the possibility that the change resulted from operator action rather than an external attacker, and the FBI did not publicly confirm a targeted intrusion. Regardless of cause, the incident exposed weaknesses in remote-access governance at small utilities.
\end{itemize}

\subsection{Comprehensive Incident Catalogue (2010--2025)}

Table~\ref{tab:full_timeline} lists 69 cross-validated OT/ICS-impacting cyber incidents from June 2010 to December 2025, sorted chronologically. Each row was verified against at least two independent sources (primary government or vendor advisory plus independent corroboration). Where full primary corroboration was not available we mark the row \emph{(single source)}; where the cited scope or attribution remains contested we mark it \emph{(disputed)}. The Type column distinguishes \textbf{D}~(direct ICS-targeting malware or attack), \textbf{R}~(ransomware on an OT-operating company), \textbf{C}~(wormable IT cascade with major OT impact), and \textbf{T}~(ICS toolkit discovered pre-deployment). A trailing question mark (e.g., \textbf{D?}, \textbf{R?}) indicates that the type assignment is itself contested, typically because the underlying incident is disputed, attribution is uncertain, or scope is contested by the named victim. The classification follows the convention of recent Dragos and Mandiant reporting, and we use it in Section~\ref{subsec:directvsindirect} to clarify the historical record.

{\scriptsize
\setlength{\tabcolsep}{3pt}
\renewcommand{\arraystretch}{1.15}
\begin{longtable}{@{}p{0.4cm}p{1.3cm}p{2.1cm}p{2.1cm}p{2.0cm}p{2.6cm}p{2.4cm}p{0.4cm}@{}}
\caption{Comprehensive cross-validated catalogue of OT/ICS-impacting cyber incidents, 2010--2025. Type: D~=~direct ICS-targeting; R~=~ransomware on OT operator; C~=~wormable IT cascade with OT impact; T~=~ICS toolkit discovered pre-deployment. Sources for each row are documented in the underlying research compendium and consist of primary government advisories (CISA, FBI, NCA, EPA, BSI, CERT-UA, NCSC), vendor reports (Dragos, Mandiant, ESET, Symantec, Claroty, SentinelOne), SEC filings, and corroborating independent reporting.}
\label{tab:full_timeline} \\
\toprule
\textbf{\#} & \textbf{Date} & \textbf{Incident} & \textbf{Sector / Geo.} & \textbf{Vector} & \textbf{OT impact} & \textbf{Scale / notes} & \textbf{T} \\
\midrule
\endfirsthead
\multicolumn{8}{c}{\tablename\ \thetable{} -- continued from previous page} \\
\toprule
\textbf{\#} & \textbf{Date} & \textbf{Incident} & \textbf{Sector / Geo.} & \textbf{Vector} & \textbf{OT impact} & \textbf{Scale / notes} & \textbf{T} \\
\midrule
\endhead
\midrule \multicolumn{8}{r}{\textit{continued on next page}} \\
\endfoot
\bottomrule
\endlastfoot
1  & 2010-06 & Stuxnet & Nuclear / Iran & USB / supply chain & PLC manipulation; centrifuge sabotage & \textasciitilde1{,}000 IR-1 centrifuges damaged & D \\
2  & 2011-02 & Night Dragon & Oil \& gas / global & Phishing, SQLi, RAT & Espionage; pre-positioning & Multi-year exfiltration & D \\
3  & 2011-10 & Duqu & Cross-sector / EU, ME & Word zero-day & Recon platform; Stuxnet code base & \textasciitilde6 confirmed orgs & D \\
4  & 2012-04 & Flame / sKyWIper & Energy, gov / Iran, ME & Multiple zero-days & Espionage; oil-terminal disconnect linkage & \textasciitilde1{,}000 infections & D \\
5  & 2012-08 & Shamoon~1 (Aramco) & Oil \& gas / Saudi Arabia & Phishing / insider & Wiper (MBR + data) & 30{,}000--35{,}000 PCs wiped & C \\
6  & 2014-06 & Havex / Dragonfly~1.0 & Energy, pharma / US, EU & Trojanized vendor installers & OPC scanning; espionage & 100s of orgs scanned & D \\
7  & 2014-12 & German Steel Mill & Steel / Germany & Spear phishing $\to$ OT & Loss of blast-furnace control & ``Massive'' physical damage & D \\
8  & 2014--17 & BlackEnergy~2/3 in US ICS & Energy, water / US & Phishing / exploit & HMI implants & Multiple US ICS HMIs & D \\
9  & 2015-12 & Ukraine Power Grid \#1 & Energy / Ukraine & BlackEnergy~3 $\to$ SCADA & Hijacked HMI; KillDisk; TDoS & \textasciitilde225k customers, 1--6\,h, 3 oblenergos & D \\
10 & 2016-12 & Industroyer / CrashOverride & Energy / Ukraine (Kyiv) & Sandworm intrusion & Native IEC-101/104, IEC-61850 abuse & Pivnichna substation; \textasciitilde1\,h, $\sim$1/5 Kyiv & D \\
11 & 2017-05 & WannaCry & Cross-sector / global & EternalBlue worm & Worm shutdown of Windows endpoints & 200k+ machines; NHS England 6{,}912 confirmed cancellations / \textasciitilde19k estimated total (NAO) & C \\
12 & 2017-06 & NotPetya & Logistics, pharma, mfg / global & M.E.Doc supply chain & Wiper masquerading as ransomware & \textasciitilde\$10B total; Maersk, Merck, FedEx & C \\
13 & 2017-08 & Triton / Trisis / HatMan & Petrochemical / Saudi Arabia & Engineering workstation $\to$ SIS & First SIS-targeted malware (Triconex) & Plant tripped to safe state & D \\
14 & 2017 & Dragonfly~2.0 & Energy / US, EU & Phishing, watering-hole & HMI screenshot capture; ICS recon & ``Hundreds'' of energy victims & D \\
15 & 2018-08 & TSMC WannaCry variant & Semiconductor / Taiwan & Unscanned tool on fab LAN & WannaCry crashed fab automation & 10k+ machines; initial est.\ \textasciitilde\$170M Q3 rev impact; booked NT\$2.6B ($\approx$\$85M) & C \\
16 & 2018-12 & Shamoon~3 / Saipem & Oil services / global & Unknown (likely creds) & RawDisk MBR overwrite & 300--400 servers & C \\
17 & 2019-03 & Norsk Hydro / LockerGoga & Aluminum / Norway, global & Phishing $\to$ AD compromise & Manual operation of smelters & 22k PCs, 170 sites; \textasciitilde\$67--84M & R \\
18 & 2019--20 & EKANS / SNAKE family & Mfg, utilities / multi & Mainly RDP & First crime-ware with ICS process kill list & Targeted ICS process termination & R \\
19 & 2020-01 & Picanol Group & Industrial machinery / EU & Unknown ransomware & All production sites encrypted & 1{,}500 workers furloughed & R \\
20 & 2020-02 & ISS World & Facility mgmt / global & Unknown & IT shutdown of global network & \textasciitilde\$74M & R \\
21 & 2020-04 & EDP / Ragnar Locker & Electric utility / Portugal & Stolen credentials & IT encryption; 10\,TB exfiltrated & €10M demand; not paid & R \\
22 & 2020-06 & Honda / SNAKE-EKANS & Auto mfg / multi & Likely RDP & EKANS targeted; multiple plants & Plants offline several days & R \\
23 & 2021-02 & Oldsmar water (\emph{disputed}) & Water / US-FL & TeamViewer (initial claim) & NaOH setpoint change (later contested) & FBI: unable to confirm intrusion & D? \\
24 & 2021-02 & Kia Motors America (\emph{disputed}) & Auto / US & DoppelPaymer (claimed) & Dealer/owner portal outage & \$20M demand; Kia denied ransomware & R? \\
25 & 2021-03 & Sierra Wireless & IoT modules / Canada & Unknown & IT encryption halted mfg & Factories down \textasciitilde1 week & R \\
26 & 2021-05 & Colonial Pipeline / DarkSide & Oil pipeline / US & Legacy VPN, no MFA & Precautionary OT shutdown & $\sim$5\,d offline (7--12 May); \$4.4M ransom (partial DOJ recovery); regional panic buying & R \\
27 & 2021-05 & Brenntag NA / DarkSide & Chemicals / US, DE & Stolen credentials & Encryption; data exfil & \$4.4M ransom paid & R \\
28 & 2021-05 & HSE Ireland / Conti & Healthcare / Ireland & Phishing; 8-week dwell & Reversion to paper across hospitals & $\approx$€102M direct cost (HSE 2024); €657M 7-yr cyber programme; €19.99M demand & R \\
29 & 2021-05 & JBS Foods / REvil & Food / US, CA, AU & Unknown & 13 plants halted & 1/5 US beef capacity; \$11M ransom & R \\
30 & 2021-09 & Olympus / BlackMatter (8 Sep) + Macaw / Evil Corp (10 Oct) & Med devices / global & Unknown & Two ransomware events $\sim$1 mo apart & EMEA IT systems + Americas IT systems disrupted & R \\
31 & 2021-11 & Volvo Cars / claimed by Snatch & Auto mfg / Sweden & File-repository unauthorized access & Limited R\&D property exfiltrated; no impact on car safety or personal data \cite{volvo_2021press} & Volvo confirms ``limited amount'' of R\&D property stolen & R \\
32 & 2022-02 & Viasat KA-SAT / AcidRain & Satcom; wind / UA, EU & VPN mgmt-plane abuse & Modem-firmware wiper & \textasciitilde5{,}800 Enercon turbines (Central Europe; $\sim$11\,GW) lost \emph{remote SCADA} & D \\
33 & 2022-02 & Toyota / Kojima Industries & Auto supply chain / Japan & Third-party partner & Tier-1 supplier IT down; JIT halted & 14 plants, 28 lines, \textasciitilde13k vehicles ($\sim$1\,day Japan output / $\sim$5\% monthly); 1-day halt & R \\
34 & 2022-02 & Bridgestone / LockBit~2.0 & Tire mfg / US, LATAM & Unknown & 30 plants disconnected & \textasciitilde10 days down & R \\
35 & 2022-03 & Nordex / Conti & Wind energy OEM / DE & Phishing $\to$ TrickBot & Remote turbine mgmt offline & Operations continued & R \\
36 & 2022-04 & Industroyer2 + CaddyWiper (foiled) & Energy / Ukraine & Sandworm intrusion & IEC-104 substation control attempt & Foiled by CERT-UA + ESET & D \\
37 & 2022-04 & INCONTROLLER / PIPEDREAM & Cross-ICS / pre-deployment & N/A (toolkit) & Schneider, OMRON, OPC-UA, CODESYS & First ICS toolkit caught pre-use & T \\
38 & 2022-05 & Foxconn Tijuana / LockBit & Electronics / Mexico & Unknown & Production disrupted & Returned to normal & R \\
39 & 2022-06 & Knauf / Black Basta & Construction / DE, global & Unknown & Email + ordering offline & Plasterboard supply impact & R \\
40 & 2022-08 & Continental / LockBit & Auto-parts / Germany & Browser drive-by & 40\,TB exfiltration; ops largely intact & \$50M demand & R \\
41 & 2022-10 & CommonSpirit Health & Healthcare / US & Unknown ransomware & EHR offline; ER diversions & 100+ facilities; 623k PHI; \textasciitilde\$160M & R \\
42 & 2022-10 & Aurubis & Copper smelting / DE, EU & Unknown & Preventive IT shutdown & Largest EU copper producer & R \\
43 & 2023-01 & Royal Mail / LockBit & Postal / UK & Unknown & Heathrow distribution centre encrypted & Intl parcel down 6 weeks; £22M & R \\
44 & 2023-02 & MKS Instruments & Semi-eq / US & Ransomware & IT + mfg disrupted (Vacuum/Photonics divisions) & \$200M Q1 hit (MKS 8-K); \$250M AMAT Q2 cascade (AMAT Q1 FY23 call) & R \\
45 & 2023-02 & Dole plc & Food / NA & Ransomware & N.\ American plants halted; \textasciitilde50\% of legacy servers + \textasciitilde25\% of end-user computers affected & \$10.5M direct cost (\$4.8M continuing-ops); 3{,}885 U.S.\ employee records breached & R \\
46 & 2023-03 & Sun Pharma / ALPHV & Pharma / India, global & Unknown & IT containment shutdown & 17\,TB claim & R \\
47 & 2023-05 & CosmicEnergy (toolkit) & Electric / pre-deployment & N/A (toolkit) & IEC-104 RTU commands & Likely red-team origin & T \\
48 & 2023-06 & Suncor / Petro-Canada & Oil retail / Canada & Unknown & Card/loyalty offline & \textasciitilde1{,}500 stations & R \\
49 & 2023-08 & Clorox & CPG / US & Unknown intrusion & IT outage cascaded into prod & \$356M Q1 FY24 net-sales decline + \textasciitilde\$49M direct cost (Clorox 8-K) & R \\
50 & 2023-10 & Boeing / LockBit & Aerospace / global & Citrix Bleed (CVE-2023-4966; CISA-confirmed) & Parts \& distribution data leak & 43--50\,GB leaked & R \\
51 & 2023-11 & CyberAv3ngers / Aliquippa & Water / US-PA, IL earlier & Internet-exposed Unitronics & PLC defacement; HMI swap & Multiple US WWS \& other sectors per CISA AA23-335A & D \\
52 & 2024-01 & FrostyGoop / Lviv & District heating / Ukraine & Internet-facing MikroTik & First Modbus-TCP-native ICS malware & \textasciitilde600 buildings, sub-zero & D \\
53 & 2024-01 & Schneider Electric / Cactus & ICS vendor / France & Unknown & Sustainability biz IT data theft & 1.5\,TB claim & R \\
54 & 2024-01 & Hyundai Motor Europe / Black Basta & Auto / EU & Unknown & IT data theft & 3\,TB claim & R \\
55 & 2024-02 & Change Healthcare / ALPHV & Healthcare clearing / US & Citrix portal w/o MFA & Cascade into pharmacy/claims & $\sim$190M (Jan 2025) / $\sim$192.7M (Jul 2025) individuals; \textasciitilde\$2.46B total & R \\
56 & 2024-04 & Fuxnet / Blackjack (Moscollector) & Sensors / Russia (Moscow) & Insider/network access & Wiper for sensor gateways & \textasciitilde500 gateways bricked (Claroty); 87k claimed by actor & D \\
57 & 2024-05 & Ascension Health / Black Basta & Healthcare / US & Employee downloaded malicious file & EHR/lab/radiology offline; ER diversions & 142 hospitals (19 states + DC); 5{,}599{,}699 records; \$1.8B FY24 op.\ loss (partial cyber attribution) & R \\
58 & 2024-06 & Synnovis / Qilin (NHS pathology) & Healthcare lab / UK-London & Unknown & Pathology lab encrypted & 7 hospitals; 1{,}134 ops cancelled & R \\
59 & 2024-06 & CDK Global / BlackSuit & Auto retail (DMS) / US, CA & Unknown & Two-stage SaaS encryption & \textasciitilde15k dealerships; \textasciitilde\$1.02B AEG-modelled dealer losses & R \\
60 & 2024-08 & Halliburton / RansomHub & Oil services / global & Phishing (reported) & IT systems disrupted; client-system disconnections & $\sim$\$35M direct cost disclosed (Halliburton 8-K) & R \\
61 & 2024-08 & Stoli Group & Beverages / US & Unknown & ERP and accounting forced manual & Cited as contributory factor in 27 Nov 2024 Ch.\ 11 filing (Tex.\ Bankr.\ N.D.\ \#24-80146-swe11; converted to Ch.\ 7 Jan 2026) & R \\
62 & 2024-09 & VW / 8Base (\emph{scope disputed}) & Auto / Germany & Unknown & Data extortion claim & VW disputes scope & R? \\
63 & 2024-11 & Schneider Electric / Hellcat & ICS vendor (Jira) / France & Unknown & Internal Jira platform breach & 40\,GB; ransom ``in baguettes'' & R \\
64 & 2025-01 & Tata Technologies / Hunters & Eng services / India & Unknown & IT services suspended & 1.4\,TB claim & R \\
65 & 2025--26 & Iran CyberAv3ngers vs US PLCs (CISA AA26-097A) & Water, energy, govt / US & Internet-exposed Rockwell A-B PLCs & PLC project-file mods; HMI/SCADA data manipulation & Multi-sector disruption; some op./financial loss & D \\
66 & 2025-04 & M\&S + Co-op + Harrods & Retail / UK & Helpdesk SE; AD takeover & Encryption; online halt; SAP outage & Combined M\&S+Co-op £270--440M (CMC Cat.\ 2); M\&S \textasciitilde£300M profit hit; 4 arrests Jul 2025 & R \\
67 & 2025-08 & Pakistan Petroleum / Blue Locker & Oil \& gas / Pakistan & Unknown & IT \& financial systems encrypted; core ops unaffected & 2-day IT/financial outage; PPL reports no critical-data compromise & R \\
68 & 2025-08 & Jaguar Land Rover & Auto mfg / UK & Social engineering (claimed) & All UK production paused & $\sim$5\,wk halt; CMC-modelled \textasciitilde£108M/wk UK-mfg loss; \textasciitilde5{,}000 vehicles/wk; £1.9B UK economy modelled & R \\
69 & 2025-12 & Polish renewables (CERT Polska) & Wind/solar/CHP / Poland & Edge-device exploitation & Wiper; HMI data destruction; RTU firmware corruption; loss of view/control & $>$30 wind/PV farms + 1 CHP plant; no generation impact & D \\
\end{longtable}
}

\subsection{Direct ICS-Targeting versus Ransomware on OT Operators}
\label{subsec:directvsindirect}

A common confusion in the OT-security literature is to treat any cyberattack on a company that operates OT as an ``OT cyberattack.'' The 69-incident catalogue makes the distinction empirically tractable.

\textbf{Direct ICS-targeting attacks.} The subset of incidents in which adversaries explicitly engaged ICS protocols, controllers, safety systems, or process variables consists of: Stuxnet, Night Dragon (espionage on operational data), Duqu, Flame, Havex/Dragonfly~1.0, German Steel Mill, BlackEnergy in U.S. ICS, Ukraine Power Grid 2015, Industroyer (2016), Triton/Trisis (2017), Dragonfly~2.0, Industroyer2 (2022, foiled), CyberAv3ngers/Aliquippa (2023), FrostyGoop (2024), Fuxnet (2024), and the 2025 Iran-affiliated PLC campaign. INCONTROLLER and CosmicEnergy belong to a related but distinct category: \emph{ICS toolkits discovered before deployment}.

\textbf{Ransomware on OT operators.} The largest subset of the catalogue---categorised as Type~R in Table~\ref{tab:full_timeline}---consists of conventional ransomware operations against companies that happen to operate OT (Norsk Hydro, JBS, Colonial Pipeline, HSE, CommonSpirit, Ascension, Synnovis, CDK, Halliburton, Clorox, Royal Mail, MKS, Dole, Aurubis, Bridgestone, Toyota/Kojima, Continental, Knauf, Boeing, Suncor, Olympus, JLR, M\&S, Tata Technologies, Sun Pharma, Nordex, EDP, ISS, Picanol, Honda, Foxconn, Sierra Wireless, Stoli, Brenntag, Volvo, Hyundai Europe, Schneider Electric, and others). In these cases the OT itself was rarely directly compromised; the operational impact came from precautionary shutdowns, IT-system dependency on encrypted hosts, or just-in-time supply-chain coupling to compromised IT.

\textbf{Wormable IT cascades.} A small but historically significant set---WannaCry, NotPetya, TSMC, Shamoon~1 and~3---consists of self-propagating IT-side malware whose worm behavior reached OT-adjacent systems through SMB, file-share, or fab-LAN propagation, producing OT impact without ICS-specific design.

The distinction matters because the defensive priorities differ. For \emph{direct ICS-targeting}, investment in ICS protocol-aware monitoring, signed firmware, network segmentation at the OT-protocol layer, and OT-specific threat intelligence yields the highest marginal return. For \emph{ransomware on OT operators}, the dominant risk-reduction levers are conventional IT-security hygiene (MFA, identity-centric access, patching, supply-chain assurance) plus segmentation that prevents IT compromise from forcing precautionary OT shutdowns. The 2024--2025 record shows both attack patterns expanding simultaneously, and a balanced investment posture is required.

\subsection{Commercial Effects by Sector}
\label{subsec:commercial_effects}

The aggregate-level financial picture from cross-survey reporting is straightforward: approximately half of surveyed organizations experienced an OT-impacting intrusion in the prior twelve months \cite{fortinet2024ot}; the August 2025 Dragos--Marsh McLennan \emph{OT Security Financial Risk Report} models a severe-but-plausible 1-in-250-year tail-event scenario at US\$329.5 billion in global losses, with approximately US\$172.4 billion attributed to OT-related business interruption (the same study places average annual OT-related cyber risk at approximately US\$31.1 billion) \cite{dragos2025risk}; and roughly a quarter of industrial firms have sustained per-incident damages exceeding US\$5 million \cite{kaspersky2024industrial}. These are useful order-of-magnitude benchmarks but they obscure substantial sectoral variation. The remainder of this subsection grounds the picture in concrete, source-traceable cost figures from individual incidents in the catalogue (Table~\ref{tab:full_timeline}). For each major sector we cite incidents whose financial impact appears in primary disclosure (SEC 10-K and 8-K filings, government post-incident reviews, regulator notifications, public press releases) and identify the dominant cost driver---ransom paid, restoration cost, lost revenue or production, or cascading supplier impact.

Table~\ref{tab:sector_concrete} consolidates the per-sector breakdown that follows.

{\scriptsize
\renewcommand{\arraystretch}{1.25}
\setlength{\tabcolsep}{4pt}
\begin{longtable}{|p{2.5cm}|c|p{4.4cm}|p{4.5cm}|p{2.4cm}|}
\caption{Sector-by-sector commercial impact of OT/IT cyber incidents, anchored to concrete disclosed figures from primary sources. ``Catalogue n'' is the count of incidents in this sector from Table~\ref{tab:full_timeline}.}
\label{tab:sector_concrete} \\
\hline
\textbf{Sector} & \textbf{Cat.\ n} & \textbf{Notable disclosed financial impact} & \textbf{Dominant cost drivers} & \textbf{Primary sources} \\ \hline
\endfirsthead
\multicolumn{5}{c}{\tablename\ \thetable{} -- continued from previous page} \\
\hline
\textbf{Sector} & \textbf{Cat.\ n} & \textbf{Notable disclosed financial impact} & \textbf{Dominant cost drivers} & \textbf{Primary sources} \\ \hline
\endhead
\hline \multicolumn{5}{r}{\textit{continued on next page}} \\
\endfoot
\hline
\endlastfoot
Oil, gas, and pipelines & 7 & Colonial Pipeline (2021): \$4.4M ransom + \textgreater\$25M/day industry-estimated operating loss across $\sim$5-day pipeline outage; Halliburton (2024): \textasciitilde\$35M direct cost (8-K); Suncor (2023): \textasciitilde1{,}500 stations on cash-only & Multi-day shutdowns; payment-system disruption; supply-chain panic buying & \cite{guardian2021colonial, cisa2023colonial} (Colonial); \cite{halliburton_8k_aug2024} (Halliburton 8-K); Suncor press releases \cite{suncor_2023press} \\ \hline
Electric power and grid & 7 & Ukraine 2015: \textasciitilde225k customers, 1--6\,h; Industroyer (2016): \textasciitilde1\,h Kyiv; AcidRain (2022): \textasciitilde5{,}800 Enercon wind turbines (Central Europe; $\sim$11\,GW) lost remote SCADA; EDP/Ragnar Locker (2020): €10M demand (refused) & Direct ICS protocol abuse; precautionary outages; loss of remote control & CISA IR-ALERT-H-16-056-01; ESET; SentinelLabs; Enercon press statement \\ \hline
Water and wastewater & 4 & Aliquippa (2023): MWAA booster station for Raccoon/Potter Twp.\ on manual fallback (\textasciitilde6{,}600 customer accounts in MWAA system); Iran-PLC 2025: Irish utility \textasciitilde2 days no water; Oldsmar (2021, disputed) & Defaced HMIs on internet-exposed PLCs; manual-fallback duration & CISA AA23-335A \cite{cyberavengers2023}; EPA/FBI/CISA/NSA 2025 \cite{iranplc2026} \\ \hline
Healthcare and biomedical & 6 & Change Healthcare (2024): \textasciitilde\$2.46B total cost, \$22M ransom, $\sim$190M individuals impacted (revised $\sim$192.7M, Jul 2025); Ascension (2024): \$1.8B FY24 operating loss; HSE Ireland (2021): \textasciitilde€102M direct cost (HSE, 2024); CommonSpirit (2022): \textasciitilde\$160M; Synnovis (2024): 1{,}134 ops cancelled in 13\,d & Restoration cost; clinical-service disruption; revenue loss; PHI breach notification & UnitedHealth Group 8-K \cite{changehc2024}; Ascension \cite{ascension2024}; HSE PIR \cite{hsepostreview, hse2024cag}; CommonSpirit 10-K \cite{commonspirit10k}; NHS England \cite{synnovis2024} \\ \hline
Automotive manufacturing & 9 & Jaguar Land Rover (2025): \textasciitilde£108M/wk × 5\,wk modelled UK-mfg loss, £1.9B UK economy modelled; CDK Global (2024): \textasciitilde\$1.02B AEG-modelled dealer losses + \$25M ransom; Toyota/Kojima (2022): 14 plants, \textasciitilde13k vehicles ($\sim$1 day Japan output) lost; Norsk Hydro (2019): \$67--84M; Continental (2022): \$50M demand; Honda (2020): plants down several days; Bridgestone (2022): 30 plants \textasciitilde10\,d & Just-in-time supply-chain amplification; SaaS-platform compromise affecting thousands of dealers; manual-fallback feasibility varies & UK Cyber Monitoring Centre \cite{ cmc_jlr2025}; CNN/Cybersecurity Dive \cite{cdkglobal2024}; Toyota \cite{toyotakojima2022, computerweekly_toyota2022}; Microsoft \cite{norskhydro2019} \\ \hline
Chemicals, materials, copper, semiconductor & 8 & Saint-Gobain (NotPetya, 2017): \$384M; TSMC (2018): \textasciitilde\$170M Q3 est.\ / NT\$2.6B booked; MKS Instruments (2023): \textasciitilde\$200M Q1 (8-K) + Applied Materials \$250M Q2 cascade (AMAT call); Brenntag (2021): \$4.4M ransom paid; Aurubis (2022): largest EU copper producer disrupted; Picanol (2020): \textasciitilde1\,wk down & Self-propagating worm hitting fab automation; production halts; cascading semi-equipment supply chain & UK NCSC NotPetya attribution \cite{notpetya2018wh}; TSMC \cite{tsmc2018, tsmc2018booking}; MKS 8-K \cite{mks_8k_feb2023}; AMAT Q1 FY23 call \cite{amat_q1fy23} \\ \hline
Food, beverage, retail, CPG & 7 & JBS (2021): \$11M ransom, 1/5 U.S.\ beef capacity; Mondelez (NotPetya, 2017): \$150--188M; Clorox (2023): \$356M Q1 FY24 net-sales decline + \textasciitilde\$49M direct cost (8-K); M\&S+Co-op (2025): \textasciitilde£270--440M combined CMC; Stoli Group (2024): cited as contributory factor in Ch.\ 11 filing; Dole (2023): \$10.5M direct cost (6-K + 10-K), N.\ American plant halts & Manufacturing/distribution halts; reputational and customer-facing online disruption; ERP-driven going-concern impact & FBI \cite{jbs2021}; UK NCSC; Clorox 8-K \cite{clorox_8k_aug2023}; CMC \cite{cmc_msccoop2025}; Stoli Ch.\ 11 \cite{stoli_ch11}; Dole 6-K \cite{dole_6k_2023} \\ \hline
Postal, logistics, shipping & 4 & Maersk (NotPetya, 2017): \$200--300M (Q3 rev impact, Maersk Q2 trading update); FedEx/TNT Express (NotPetya): \$400M; Royal Mail (2023): £22M revenue loss + £10M remediation, 6-week intl outage & Container terminals reverting to manual; international parcel-service halt; high-volume operational backlog & UK NCSC NotPetya \cite{notpetya2018wh}; Royal Mail/UK Parliament BEIS \\ \hline
Pharmaceutical & 4 & Merck (NotPetya, 2017): \$870M; Sun Pharma (2023): 17\,TB claim; Olympus (2021): EMEA IT (BlackMatter, 8 Sep) and Americas IT (Macaw / Evil Corp, 10 Oct) disrupted within $\sim$1 month \cite{olympus2021emea, olympus2021americas}; Reckitt (NotPetya): not separately disclosed in detail & Cross-sector worm cascade; production-line and R\&D disruption & UK NCSC NotPetya \cite{notpetya2018wh}; SEC filings \\ \hline
Aerospace and defense & 1 & Boeing (2023, LockBit): 43--50\,GB leaked; Boeing states aircraft safety unaffected & Data-leak risk to parts and distribution unit & CISA + Boeing public confirmation \\ \hline
Cross-sector NotPetya cascade (2017) & --- & Total: \textasciitilde\$10B globally (NCSC). Component victims: Maersk \$200--300M (Q2-trading-update guidance), Merck \$870M, FedEx \$400M, Saint-Gobain \$384M, Mondelez \$150--188M, plus Reckitt and others & Single wormable wiper produced largest publicly documented multi-sector cyber loss in history & UK NCSC; White House 2018 attribution \cite{notpetya2018wh} \\ \hline
\end{longtable}
}

\subsubsection{Oil, Gas, and Pipelines}
Oil-and-gas losses are driven overwhelmingly by precautionary multi-day shutdowns of high-throughput infrastructure, with ransom payments negligible against revenue lost. Colonial Pipeline is the canonical benchmark: a \$4.4M ransom was trivial beside the regional fuel disruption from a five-day precautionary shutdown of a system carrying \textasciitilde45\% of U.S.\ East Coast fuel \cite{guardian2021colonial, cisa2023colonial}. The same pattern---modest direct cost, large operational impact---recurs in Halliburton (2024) \cite{halliburton_8k_aug2024} and Suncor (2023), whose incident forced cash-only operation across \textasciitilde1{,}500 retail stations \cite{suncor_2023press}.

\subsubsection{Electric Power and Grid}
Power-grid incidents produce comparatively small direct financial losses but disproportionate strategic significance; the dominant cost-and-risk driver is regulatory and reputational rather than monetary. Outages of the documented durations (Ukraine 2015, Industroyer 2016) were operationally manageable through manual breaker operation yet politically consequential. AcidRain (2022) illustrates the sector's distinctive failure mode: it severed remote SCADA control of \textasciitilde5{,}800 Enercon turbines without stopping generation---the turbines kept spinning, but operators lost the ability to adjust them \cite{acidrain2022, enercon2022}. EDP's refusal of a €10M Ragnar Locker demand (2020) underscores that ransom is rarely the material cost here.

\subsubsection{Water and Wastewater}
Water-sector incidents carry small per-incident financial impact but high public-safety and political salience, and characteristically exploit a low technical bar---internet-exposed PLCs running default credentials rather than sophisticated tradecraft. Aliquippa (2023) placed a booster station on manual fallback without water-quality impact \cite{cyberavengers2023}, and the 2025--2026 Iran-affiliated campaign extended the same pattern to Rockwell Allen-Bradley controllers across multiple U.S.\ utilities \cite{iranplc2026}. The dominant cost driver is regulatory response and reputational damage; per-incident losses are low, but the political pressure for federal water-sector cybersecurity mandates has been substantial---amplified by the disputed Oldsmar (2021) event, which the FBI could not confirm as an intrusion.

\subsubsection{Healthcare and Biomedical}
Healthcare exhibits the highest per-incident financial impact of any sector, dominated by restoration cost, clinical-service-revenue loss, and PHI breach-notification compliance rather than ransom (typically a small fraction of exposure). The scale is set by Change Healthcare (2024), whose \textasciitilde\$2.46B total cost and \textasciitilde190M-individual breach rank it among the costliest single events in U.S.\ cyber history \cite{changehc2024, hhs_changehc_faq}, and Ascension (2024), which reported a \$1.8B FY24 operating loss \cite{ascension2024}. HSE Ireland \cite{hsepostreview, hse2024cag}, CommonSpirit \cite{commonspirit10k}, and Synnovis \cite{synnovis2024} confirm the pattern; Section~\ref{subsec:healthcare} develops the healthcare threat model in depth.

\subsubsection{Automotive Manufacturing}
Automotive demonstrates the just-in-time supply-chain amplification effect more clearly than any other sector: compromise of a single supplier or platform halts the largest manufacturers for days or weeks. Jaguar Land Rover (2025) is the extreme case---a five-week UK production stoppage with a modelled \textasciitilde£1.9B UK-economy impact through supplier ripple, the largest such figure to date \cite{cmc_jlr2025}. CDK Global (2024) generalised the effect to the SaaS layer, reverting \textasciitilde15{,}000 dealerships to paper for three weeks (\textasciitilde\$1.02B modelled dealer losses) \cite{cdkglobal2024, aeg2024cdk}, and Toyota/Kojima (2022) showed a tier-1 supplier's IT compromise idling 14 Toyota plants in a day \cite{toyotakojima2022, computerweekly_toyota2022}. Manual-fallback feasibility varies: Norsk Hydro kept smelters running by hand \cite{hydro_ar2019, hydro_cyberpage}, whereas JIT assembly has no such cushion.

\subsubsection{Chemicals, Materials, Copper, and Semiconductor}
This sub-sector's financial profile is defined by wormable IT cascades reaching fab-automation and plant-DCS hosts that share Windows infrastructure with corporate IT. TSMC's 2018 WannaCry-variant infection of \textasciitilde10{,}000 fab hosts \cite{tsmc2018, tsmc2018booking} and NotPetya's hit on Saint-Gobain \cite{notpetya2018wh} exemplify the mechanism. The 2023 MKS Instruments ransomware adds a second pattern---equipment-supply-chain coupling---its \textasciitilde\$200M revenue impact \cite{mks_8k_feb2023} cascading into an \textasciitilde\$250M hit at customer Applied Materials \cite{amat_q1fy23}. Precautionary shutdowns at Brenntag, Aurubis, and Picanol round out the picture.

\subsubsection{Food, Beverage, Retail, and Consumer Packaged Goods}
In food, beverage, and retail the dominant driver is extended IT-system unavailability cascading into manufacturing and distribution, amplified by unusually high public visibility (panic buying, supermarket shortages). JBS (2021) halted \textasciitilde1/5 of U.S.\ beef capacity \cite{jbs2021}; Clorox (2023) reported a \textasciitilde\$356M net-sales decline \cite{clorox_8k_aug2023}; and the M\&S/Co-op attacks (2025) disrupted food distribution with a combined \textasciitilde£270--440M modelled impact \cite{ms_coop_2025, cmc_msccoop2025}. The sector also furnishes the clearest going-concern case: Dole's plant halts \cite{dole_6k_2023} and Stoli Group's 2024 ransomware---cited as a contributory factor in its subsequent Chapter~11 filing \cite{stoli_ch11}.

\subsubsection{Postal, Logistics, and Shipping}
Logistics is defined by the high cost of even short downtime in operations whose throughput depends on real-time scheduling and tracking. Maersk's NotPetya loss (\textasciitilde\$200--300M) remains the canonical reference---17 of 76 APM terminals reverted to manual paper dispatch and 45{,}000 PCs were rebuilt over ten days \cite{maersk2017q2update, maritimeexec_maersk2017}. FedEx/TNT absorbed \textasciitilde\$400M from the same cascade, and Royal Mail's 2023 LockBit incident took international parcel service offline for \textasciitilde6 weeks.

\subsubsection{The NotPetya Cascade in Context}
The 2017 NotPetya event remains the largest publicly documented cross-sector cyber loss---\textasciitilde\$10B globally per UK NCSC \cite{notpetya2018wh}, spanning Maersk, Merck (\$870M), FedEx, Saint-Gobain, and Mondelez. A single wormable wiper distributed through a compromised accounting-software update establishes the empirical upper bound on what one supply-chain compromise can inflict across IT-coupled OT operators, and it remains the principal anchor for high-end OT risk modelling.

A consistent qualitative pattern across the sector subsubsections above is that the breakdown of OT-incident cost typically apportions the dominant share to production loss and restoration, a smaller share to revenue loss from disrupted services and customers, and a comparatively small share to ransom payments themselves. Across Colonial Pipeline (\$4.4M ransom against \$25M-plus daily revenue loss), JBS (\$11M ransom in a multi-plant production halt), Norsk Hydro (no ransom paid; \$67--84M total losses), and CDK Global (\$25M ransom against \textasciitilde\$1B industry impact), the ransom-to-total-cost ratio is typically below 5\% and often below 1\%. We do not display this as an aggregate pie chart because the underlying victim disclosures are not categorized on a common cost-component schema.
\begin{figure}[h]
    \centering
    \includegraphics[width=1.0\linewidth]{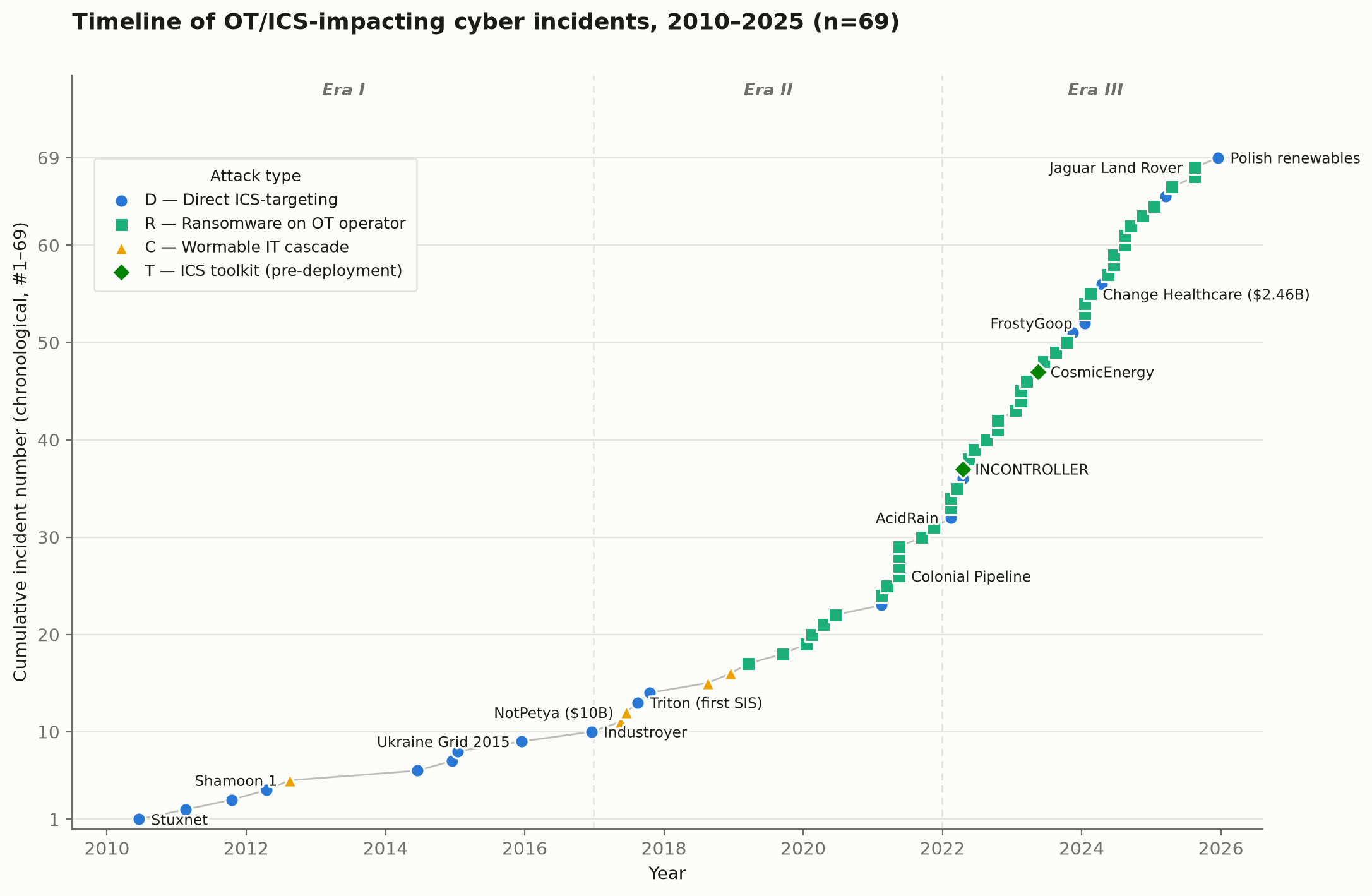}
    \caption{Timeline of notable attacks- and ICS-targeted attacks,  A comprehensive 69-incident catalogue spanning 2010--2025 is provided in Table~\ref{tab:full_timeline}.}
    \label{fig:attack_timeline}
\end{figure}

Figure~\ref{fig:attack_timeline} plots the 69 catalogued incidents as a cumulative
curve, with each point's colour and shape encoding its Type. The curve is nearly flat
through Era~I (2010--2016): fewer than two incidents per year, and almost every point
is a blue \emph{direct ICS-targeting} marker---the era of bespoke, high-effort
operations such as Stuxnet, Ukraine~2015, and Industroyer. From 2021 onward the slope
steepens sharply, and the acceleration is dominated by green \emph{ransomware-on-OT-operator}
squares (43 of the 69 incidents, versus 19 direct-ICS, 5 wormable cascades, and 2
pre-deployment toolkits). In other words, the recent growth in OT-impacting incidents
is driven far less by advances in ICS-specific malware than by conventional ransomware
reaching companies that happen to operate OT---the distinction developed in
Section~\ref{subsec:directvsindirect}. The two ICS-toolkit markers (INCONTROLLER,
CosmicEnergy) sit apart as capabilities caught \emph{before} deployment rather than
realised incidents.

\subsection{Mitigation Effectiveness: A Qualitative View}
Cross-survey data also indicates the relative impact of emerging defensive technologies, though no single survey instruments all four technologies on identical samples. Table~\ref{tab:mitigation} summarizes the qualitative pattern observed across 2024--2025 industry surveys (Fortinet \cite{fortinet2024ot}, Claroty \cite{claroty2024healthcare}, and Dragos/Marsh modelling). We report adoption, resilience gain, and detection-latency reduction as ordinal bands (low, medium, high) rather than as percentages, because the underlying surveys differ in sample frame and instrumentation. Treating the bands as exact percentages would overstate their precision.

\begin{table}[htbp]
\centering
\footnotesize
\caption{Indicative ordinal pattern of adoption and resilience characteristics of emerging OT defensive technologies as reported across 2024--2025 industry surveys (Fortinet \cite{fortinet2024ot}, Claroty \cite{claroty2024healthcare}). Bands are L (low), M (medium), H (high); they indicate cross-survey consensus rather than benchmark measurements.}
\label{tab:mitigation}
\renewcommand{\arraystretch}{1.2}
\begin{tabularx}{0.98\linewidth}{@{} l X c c c @{}}
\toprule
\textbf{Technology} &
\textbf{Primary objective} &
\makecell{\textbf{Reported}\\\textbf{adoption}} &
\makecell{\textbf{Reported}\\\textbf{resilience gain}} &
\makecell{\textbf{Detection}\\\textbf{latency red.}} \\
\midrule
AI/ML anomaly detection & Early intrusion identification & M & M & H \\
Zero Trust Architecture & Identity-centric access control & H & H & H \\
Blockchain-based logging & Tamper-proof event/firmware records & L & L--M & N/A \\
Digital twins & Virtual attack simulation, proactive defense & L & M & L--M \\
\bottomrule
\end{tabularx}
\end{table}

The qualitative pattern across surveys is consistent: AI/ML-based detection delivers the largest reductions in time-to-detect; ZTA delivers the largest reductions in lateral propagation and time-to-contain; blockchain and digital twins are at earlier adoption stages but provide complementary capabilities (tamper-evidence and rehearsal, respectively). Detection-latency-reduction figures specifically should not be aggregated across surveys without controlling for sample frame, and we deliberately do not display them as numeric bars.

\subsection{Sectoral Deep-Dive: case study of Healthcare OT}
\label{subsec:healthcare}

Healthcare merits an extended treatment because it concentrates the OT/IT-convergence threat model under conditions where disruption affects life safety, not just economics. Disruption to clinical or facility OT has direct patient-safety consequences---ambulance diversions, postponed surgeries, manual workarounds for infusion and ventilation---that have no analog in conventional IT breaches. Healthcare is also the sector where AI-driven defenses are showing the earliest measurable returns and where standards alignment (IEC~80001-1, FDA, NIST SP~1800-8) is most active. Recent sector studies consistently rank healthcare among the most-targeted sectors and place per-incident cost at the top of the cross-sector ranking \cite{ibm2025xforce, claroty2024healthcare}.

\subsubsection{Three Interlocking Layers}
A modern healthcare OT stack comprises three layers whose interactions define the attack surface.

\begin{enumerate}
    \item \textbf{Clinical IT.} Electronic Health Records (EHR), Picture Archiving and Communication Systems (PACS), Laboratory Information Systems (LIS), and administrative platforms. Compromise of clinical IT typically halts admissions, scheduling, billing, and clinical decision support, and forces a reversion to paper workflows.
    \item \textbf{IoMT (also termed Internet of Healthcare Things).} Network-connected medical devices: infusion pumps, ventilators, anesthesia workstations, imaging suites (CT, MR, ultrasound), patient monitors, dialysis machines. Many devices ship with limited authentication, infrequent firmware updates, and 10--20-year deployed lifecycles, mirroring OT outside healthcare.
    \item \textbf{Facility OT and Building Management Systems (BMS).} SCADA-style control of HVAC, medical-gas distribution, water purification, sterilization (autoclaves), elevators, fire-detection, and physical access. Disruption of facility OT can render operating rooms unusable or compromise sterile-processing pipelines.
\end{enumerate}

Weak segmentation between these layers is the principal architectural risk. Industry surveys report that flat or poorly segmented networks persist in a substantial minority of hospitals, exposing IoMT and facility OT to IT-borne threats such as ransomware that originate in clinical IT or administrative systems \cite{claroty2024healthcare}. Applying Zero-Trust segmentation and traffic-aware zoning informed by analytics measurably reduces cross-segment exposure and accelerates containment \cite{ibm2025xforce, nist1800-8}.

\subsubsection{Threat Landscape and Recent Trends}
Healthcare consistently ranks among the most-targeted sectors globally; IBM places average breach cost at the highest of any sector for several consecutive years \cite{ibm2025xforce}. Claroty's 2024 CPS/OT survey indicates that the majority of healthcare organizations experienced at least one OT/IoMT-impacting incident in the prior 12 months, with a substantial share reporting ransom demands above US\$1 million \cite{claroty2024healthcare}. Initial access typically comes through IT-side vectors (phishing, credential theft, vulnerable remote access) followed by lateral movement into IoMT or BMS. The HSE Ireland (2021) and CommonSpirit Health (2022) incidents \cite{hsepostreview, commonspirit10k} demonstrate that restoration and lost-revenue costs dominate financial exposure, running into the hundreds of millions of dollars in the largest events.

\subsubsection{Where AI Adds Value in Healthcare OT}
Four AI/ML application areas have shown earliest practical returns in healthcare OT.

\begin{itemize}
    \item \textbf{IoMT and device-network defense.} Unsupervised and semi-supervised models learn normal device communication patterns (protocol mix, command cadence, data volumes) and flag deviations such as command-frequency bursts, atypical data exfiltration volumes, or scan-time anomalies---reducing detection latency and improving triage in environments where IoMT devices number in the thousands per hospital \cite{claroty2024healthcare}.
    \item \textbf{Facility OT and PLC-telemetry analytics.} Predictive models on SCADA/PLC signals from BMS detect control-loop drift, actuator irregularities, or abnormal scan times that may signal compromise or impending equipment failure, supporting both cybersecurity and predictive maintenance \cite{fortinet2024ot}.
    \item \textbf{Risk and ROI dashboards for executives.} Translating ML-detected anomalies and incident-history data into Annual Risk Exposure (ARE $=\lambda\times \mathrm{SLE}$, where $\lambda$ is the annualized rate of
   occurrence), with $\mathrm{SLE}$ components tied to downtime, lost revenue, and quality loss, supports board-level cybersecurity-investment decisions in an environment where cybersecurity competes for capital with clinical priorities.
    \item \textbf{Segmentation optimization.} AI-driven traffic clustering identifies natural communication groups across IT, IoMT, and BMS, informing Zero-Trust zoning and micro-segmentation policies that minimize lateral paths from corporate IT to clinical-impact systems \cite{ibm2025xforce, nist1800-8}.
\end{itemize}

\subsubsection{Standards and Compliance Posture}
Three documents anchor healthcare OT compliance, all detailed in Section~\ref{sec:reg_frameworks}: IEC~80001-1:2021 \cite{iec80001} for risk management of IT networks incorporating medical devices; the U.S. FDA's 2023 \emph{Cybersecurity in Medical Devices} guidance \cite{fda2023} for premarket and post-market manufacturer expectations including SBOMs and vulnerability handling; and NIST SP~1800-8 \cite{nist1800-8} as a reference architecture for securing wireless infusion pumps and analogous IoMT classes. Mapping IEC~62443 controls to IEC~80001-1 processes provides a unified cross-walk for hospitals and device manufacturers seeking to harmonize control baselines.

\subsubsection{Implementation Roadmap}
A practical four-phase plan to operationalize AI-augmented defense in healthcare OT, derived from sector guidance and incident-analysis lessons:

\begin{enumerate}
    \item \textbf{Visibility and telemetry.} Inventory IoMT and facility-OT assets; instrument network choke points and PLC/BMS gateways; centralize telemetry into an OT-aware SIEM with sufficient retention for forensics.
    \item \textbf{Behavioral baselines.} Train unsupervised or semi-supervised ML models on observed device-by-device behavior; tune to acceptable false-positive rates for clinical environments where alert fatigue itself is a patient-safety concern.
    \item \textbf{Automated response.} Tie detections to micro-segmentation policies and SOAR playbooks that can isolate suspicious flows in under one second, with explicit safe-state behaviors for clinically critical IoMT (e.g., never severing a network connection that interrupts active patient care).
    \item \textbf{Continuous validation.} Use tabletop exercises and OT-lab or digital-twin rehearsals against ransomware and supply-chain scenarios; refine playbooks; track Mean Time to Detect (MTTD), Mean Time to Recover (MTTR), Containment Time, and the Operational Recovery Index (ORI) tied to clinical-service throughput.
\end{enumerate}

Industry reporting indicates a consistent qualitative pattern in healthcare OT: as AI-driven detection and segmentation maturity increases, both incident downtime and per-incident financial impact decrease; the relationship is documented at the survey level by Claroty \cite{claroty2024healthcare} and IBM X-Force \cite{ibm2025xforce}, though no public dataset cross-tabulates AI maturity against per-incident outcomes on a single sample frame. We therefore do not reproduce a numeric cross-tabulation here; the directional pattern is consistent with the concrete healthcare-incident financial impact summarised in Section~\ref{subsec:commercial_effects} (Change Healthcare, Ascension, CommonSpirit, HSE, Synnovis).

\subsection{Cross-Cutting Lessons}
Five lessons emerge consistently across the 69-incident historical record:

\begin{enumerate}
    \item \textbf{IT-side initial access drives OT-side impact.} The dominant initial-access vectors (phishing, exposed remote access, supply-chain compromise) are not OT-specific. Across the catalogue, more than two-thirds of incidents began on the IT side and produced OT consequences through precautionary shutdown or IT/OT coupling rather than through direct ICS exploitation. Defending the IT--OT boundary is therefore mostly an IT-security problem, but with OT-specific consequences.
    \item \textbf{Business interruption dominates breach cost.} Across Colonial Pipeline, Norsk Hydro, Maersk (NotPetya), CommonSpirit, Change Healthcare, CDK Global, JLR, and HSE Ireland, restoration and downtime costs eclipse ransom payments and direct data-loss exposure---typically by an order of magnitude. Change Healthcare's \textasciitilde\$2.46\,B and JLR's £1.9\,B economy-wide impact illustrate the scale. Resilience investments that reduce time-to-contain and time-to-recover yield the highest marginal financial benefit.
    \item \textbf{Adversary capability is consolidating, then democratizing.} Stuxnet was bespoke and required nation-state engineering. INCONTROLLER (2022) showed reusable, vendor-agnostic ICS post-exploitation tooling. CyberAv3ngers (2023) and the 2025 Iran-PLC campaign show the opposite end of the spectrum: opportunistic exploitation of internet-exposed Unitronics and Rockwell PLCs running default credentials. The capability bar has both risen (toolkits) and fallen (default-credential PLC scanning).
    \item \textbf{Just-in-time supply chains amplify ransomware impact.} Toyota/Kojima (2022) and Jaguar Land Rover (2025) show that cyberattacks on a single supplier or platform can halt the largest manufacturers in a sector for days or weeks. CDK Global (2024) generalized the pattern to SaaS providers serving thousands of downstream OT operators. The single-supplier bottleneck is now a primary lens for cyber-physical risk modelling.
    \item \textbf{Wartime cyber operations now routinely target ICS.} The 2022--2025 Ukraine--Russia conflict produced AcidRain, Industroyer2, FrostyGoop, and Fuxnet---a steady cadence of wiper and ICS-protocol-native malware against civilian infrastructure on both sides. Industroyer2 was the first publicly documented near-miss ICS attack during active armed conflict; FrostyGoop produced direct civilian-welfare harm. The wartime-ICS-operations precedent is now firmly established and is likely to inform future state-on-state planning.
\end{enumerate}

These lessons motivate the open-problems discussion in Section~\ref{sec:conclusion}.

\section{Regulatory Frameworks and Compliance}
\label{sec:reg_frameworks}

Regulatory and standards frameworks are the institutional layer that turns the tools and architectural choices of Section~\ref{sec:tools_gaps} into auditable, enforceable practice. Three drivers have made compliance a more central part of OT security in the last five years. First, the financial-impact data summarized in Section~\ref{sec:history_impact} has moved cybersecurity from a discretionary expense into a board-level operational-risk concern, with regulators following the money. Second, sector-specific incidents (Colonial Pipeline, the European wartime attacks on Ukrainian and adjacent infrastructure, healthcare ransomware) have triggered new mandates---NIS2, DORA, the Cyber Resilience Act---that explicitly target OT and critical-infrastructure sectors. Third, IT--OT convergence has forced a re-evaluation of frameworks designed for one or the other. This section consolidates the U.S., European, and international frameworks practitioners must navigate, and identifies how they map onto one another \cite{bellamkonda2020cybersecurity, fortunato2020risk, qiu2020edge}.

\subsection{NIST Cybersecurity Framework and NIST SP 800-82}
\label{subsec:nist}

The U.S. National Institute of Standards and Technology (NIST) maintains two pillars relevant to OT: the Cybersecurity Framework (CSF) for cross-sector risk-based program design, and Special Publication~800-82 for industrial control systems specifically.

\subsubsection{NIST CSF 2.0 (2024)}
The NIST CSF was first released in 2014 and has been revised repeatedly, culminating in CSF~2.0 (NIST CSWP 29, published 26 February 2024) \cite{nistcsf2, lanz2024updated, varol2024enhancing}. Three changes in CSF~2.0 are particularly relevant to OT.

\textbf{Govern Function added.} CSF~2.0 introduces a sixth core function---``Govern''---that sits alongside Identify, Protect, Detect, Respond, and Recover. It targets organizational-risk management, supply-chain risk, and cybersecurity strategy at the executive level. For OT operators, this formalizes the practice of integrating cybersecurity outcomes into operational and safety governance rather than leaving them to a downstream IT team.

\textbf{Supply-Chain Risk Management (SCRM).} CSF~2.0 elevates SCRM from a control category to a first-class consideration following the SolarWinds (2020) \cite{martinez2021software} and Kaseya (2021) \cite{robinson2022new} compromises and the supply-chain implications surfaced by INCONTROLLER. Practical guidance includes continuous vendor assessment, hardware and firmware attestation, and AI-driven anomaly detection on supplier-related transactions and update channels.

\textbf{Zero Trust integration.} CSF~2.0 explicitly recognizes Zero Trust Architecture principles, including:
\begin{itemize}
    \item \emph{Identity-based access controls (IBAC)}, where access is granted dynamically based on verified user and device identity \cite{almuseelem2024continuous};
    \item \emph{Multi-factor and phishing-resistant authentication} (FIDO2, certificate-based credentials) \cite{xun2025building};
    \item \emph{Policy enforcement points (PEPs)} that validate every request in real time \cite{fernandez2024critical};
    \item \emph{Micro-segmentation} that limits east--west movement and reduces blast radius.
\end{itemize}
Industry surveys consistently report rapid Zero-Trust adoption growth: Okta's 2023 \emph{State of Zero Trust Security} report places adoption at 61\% in 2023, up from a 2021 baseline of approximately 24\% \cite{OKTA}; complementary discussions appear in \cite{csoonline, syed2022zero, sample2022zta}. OT-specific deployment lags enterprise IT.

\subsubsection{NIST SP 800-82 Rev. 3 (September 2023)}
NIST SP~800-82 Rev.~3, \emph{Guide to Operational Technology (OT) Security}, was published on 28 September 2023 and is the principal U.S. guidance for securing industrial control systems \cite{nist80082rev3, o2024fiscal}. The 2023 edition reflects three substantive updates over Rev.~2 (2015).

\textbf{Network segmentation and isolation.} The revision codifies established practice around \cite{kallatsa2024strategies}: industrial demilitarized zones (IDMZs) that separate enterprise IT from OT while permitting controlled flows \cite{mazur2016defining}; unidirectional security gateways (data diodes) that enforce one-way data egress in hardware \cite{heo2016design}; and air-gapped architectures for the most critical assets \cite{guri2023air}. The Colonial Pipeline incident is repeatedly cited in the rationale for stronger segmentation between IT and OT layers.

\textbf{Behavioral anomaly detection.} Rev.~3 explicitly endorses ML-driven behavioral anomaly detection as a complement to signature-based IDS, citing the inability of signature methods to catch ICS-specific malware such as Triton (2017) \cite{myung2019ics}, Industroyer (2016) \cite{makrakis2021vulnerabilities}, and INCONTROLLER (2022) \cite{firoozjaei2022evaluation}.

\textbf{Risk and supply-chain alignment.} Rev.~3 explicitly aligns with CSF~2.0's SCRM emphasis and provides ICS-specific risk-assessment templates.

\subsection{IEC 62443: International Industrial Automation Standard}
\label{subsec:iec62443}

The IEC~62443 series, developed jointly by the International Electrotechnical Commission (IEC) and the International Society of Automation (ISA), is the international benchmark for securing Industrial Automation and Control Systems (IACS). The series has a layered structure addressing policies, system requirements, and component requirements, along with conformance assessment.

\textbf{IEC~62443-2-1:2024.} The 2024 edition specifies security-program requirements for IACS asset owners. Significant changes from the 2010 edition include reorganization around Security Program Elements (SPEs), explicit harmonization with ISO/IEC~27001 information-security management to remove duplications, and a four-level maturity model (Initial, Managed, Defined, Improving) for evaluating program elements \cite{iec1}.

\textbf{IEC~62443-3-3.} Defines system-level security requirements organized around seven foundational requirements: identification and authentication control, use control, system integrity, data confidentiality, restricted data flow, timely response to events, and resource availability \cite{ISA}. Each requirement maps to four Security Levels (SL~1 through SL~4) of increasing rigor, with SL~1 protecting against casual or coincidental violation and SL~4 protecting against intentional, well-resourced, motivated, and skilled attackers using sophisticated means.

\textbf{IEC~62443-4-1 and IEC~62443-4-2.} Specify secure-product-development lifecycle requirements for IACS component vendors and technical security capabilities for individual components (PLCs, HMIs, RTUs, network devices), respectively. These parts have become the principal vehicle through which device vendors demonstrate cybersecurity compliance to industrial customers \cite{ISA, leander2019applicability}.

\textbf{Adaptation to emerging architectures.} As IIoT, cloud-integrated SCADA, and AI-driven monitoring expand the attack surface, IEC~62443 has been progressively interpreted alongside Zero Trust Network Access principles for remote PLC and RTU configuration \cite{gottel2023qualitative}, alongside SIEM and SOAR for compliance monitoring, and with attention to 5G network slicing and dynamic policy orchestration in IIoT environments \cite{moyon2020integration, oberhofer2023market}. Blockchain-based tamper-proof logging has been proposed for IEC~62443 audit-trail requirements in highly interconnected ecosystems \cite{heinl2023standard}, though adoption remains limited.

\subsection{European Regulation: NIS2, DORA, and the Cyber Resilience Act}
\label{subsec:eu}

The European Union has issued three overlapping regulations that together govern cybersecurity for OT operators and product manufacturers.

\textbf{NIS2 Directive (Directive (EU) 2022/2555) \cite{nis2_eu}.} NIS2 was adopted in December 2022 and entered into force in January 2023; the transposition deadline for Member States was 17 October 2024, with national measures applying from 18 October 2024. NIS2 replaces the original NIS Directive and substantially expands sectoral coverage to include:
\begin{itemize}
    \item energy, transport, banking, and financial-market infrastructure;
    \item healthcare, drinking water and wastewater;
    \item digital infrastructure, public administration, and space;
    \item postal services and waste management;
    \item food production, manufacturing of medical devices, chemicals, machinery, motor vehicles;
    \item digital providers and research organizations.
\end{itemize}
Operators in scope must implement cybersecurity risk-management measures (including supply-chain security, encryption, and Zero Trust principles) and report significant incidents through a three-stage timeline (NIS2 Article 23): an early-warning notification within 24 hours of awareness, an incident notification within 72 hours, and a final report within one month. Penalties for non-compliance reach the higher of €10\,M or 2\% of worldwide annual turnover for ``essential'' entities (and €7\,M or 1.4\% for ``important'' entities).

\textbf{Digital Operational Resilience Act (Regulation (EU) 2022/2554, applies from 17 January 2025) \cite{dora_eu}.} DORA is sector-specific to financial services and the ICT third parties they rely on. Its operational-resilience-testing provisions---including threat-led penetration testing (TLPT)---and ICT third-party risk-management obligations have direct relevance for OT, particularly in payment systems, market infrastructure, and operational technology supporting financial-sector facilities.

\textbf{Cyber Resilience Act (Regulation (EU) 2024/2847, in force 10 December 2024, main obligations apply from 11 December 2027) \cite{cra_eu}.} The CRA sets horizontal cybersecurity requirements for products with digital elements placed on the EU market. The CRA categorises products into a default class plus ``important'' product classes (Annex III, including industrial automation and control systems---PLCs, DCS, CNC controllers, and SCADA) and ``critical'' product classes (Annex IV, including hardware security modules, smart-card secure elements, and similar high-assurance components). Important and critical products must satisfy essential cybersecurity requirements covering secure-by-default configuration, vulnerability handling, security-update provision, and incident reporting; conformity assessment for important products typically requires harmonised-standard self-assessment or third-party notified-body assessment, while critical products are subject to mandatory third-party assessment. Note that vulnerability- and severe-incident-reporting obligations apply earlier, from 11 September 2026, and rules on the appointment of conformity-assessment bodies apply from 11 June 2026. The CRA is the first regulation to impose binding cybersecurity obligations on PLC, RTU, and OT-software vendors at the EU border, and it interacts directly with the IEC~62443-4 product-side requirements.

\subsection{Sector-Specific Frameworks}

\textbf{NERC Critical Infrastructure Protection (CIP).} The North American Electric Reliability Corporation enforces a suite of CIP standards (CIP-002 through CIP-014) governing the cybersecurity of bulk electric system (BES) cyber assets in the U.S. and Canada. NERC CIP is one of few cybersecurity regimes with binding financial penalties and is the principal compliance driver for North American electric utilities.

\textbf{Healthcare: IEC 80001-1, FDA, NIST SP 1800-8.} IEC~80001-1:2021 \cite{iec80001} (adopted in the EU as DIN EN IEC~80001-1:2023) provides risk-management requirements for IT networks incorporating medical devices and is the spine for hospital cybersecurity governance across clinical IT, IoMT, and facility OT. The U.S. FDA's 2023 \emph{Cybersecurity in Medical Devices: Quality System Considerations and Content of Premarket Submissions} guidance \cite{fda2023} formalizes threat-modelling, software bills of materials (SBOM), and post-market vulnerability-management expectations for medical-device manufacturers. NIST SP~1800-8 \cite{nist1800-8} provides a reference architecture for securing wireless infusion pumps. These three documents together form the operating framework for healthcare OT.

\textbf{Other regimes.} Sector-specific regulators include FERC (electric/gas in the U.S.), TSA Pipeline Security Directives (post-Colonial Pipeline), the U.S. EPA in water utilities, FAA and EASA in aviation, and FINRA/EBA for financial-sector operational resilience.

\subsection{Cross-Walk and Practical Guidance}

The frameworks above are convergent in their objectives---risk-based program management, segmentation, identity-centric access, monitoring, incident response, and supply-chain assurance---but differ in scope, enforcement, and audit method. Table~\ref{tab:framework_crosswalk} summarizes the cross-walk for an asset-owner audience.

\begin{table}[htbp]
\centering
\footnotesize
\caption{Cross-walk of major OT cybersecurity frameworks for asset owners.}
\label{tab:framework_crosswalk}
\renewcommand{\arraystretch}{1.2}
\begin{tabular}{|p{3.0cm}|p{2.5cm}|p{3.5cm}|p{4.5cm}|}
\hline
\textbf{Framework} & \textbf{Geography} & \textbf{Primary scope} & \textbf{Enforcement / assessment} \\ \hline
NIST CSF 2.0 & Global (U.S.-origin) & Cross-sector risk management & Voluntary; widely adopted as cross-walk basis \\ \hline
NIST SP 800-82 Rev.~3 & Global (U.S.-origin) & ICS-specific guidance & Voluntary; mandatory for U.S. federal ICS \\ \hline
IEC 62443 series & International & IACS asset owners, integrators, vendors & Conformance assessment via accredited bodies \\ \hline
EU NIS2 & EU & Essential and important entities across many sectors & Binding; financial penalties (up to €10\,M / 2\% turnover); 24h / 72h / 1-month staged reporting \\ \hline
EU DORA & EU & Financial services and ICT third parties & Binding from Jan 2025; TLPT for systemically important firms \\ \hline
EU CRA & EU & Products with digital elements (vendor-side) & Binding; main obligations from Dec 2027; CE-marking gate \\ \hline
NERC CIP & North America (BES) & Bulk electric system cyber assets & Binding; financial penalties up to \$1M/day per violation \\ \hline
IEC 80001-1:2021 (DIN EN IEC 80001-1:2023) & International / EU & Hospital IT networks with medical devices & Voluntary; harmonizes with ISO 27001 and IEC 62443 \\ \hline
FDA 2023 Guidance & U.S. & Medical-device manufacturers (premarket) & Binding via FDA premarket review \\ \hline
TSA Pipeline SDs & U.S. & Pipeline operators & Binding post-Colonial Pipeline; specific control mandates \\ \hline
\end{tabular}
\end{table}

For practitioners, three practical observations follow from this landscape.

\textbf{Use NIST CSF 2.0 as the cross-walk anchor.} Most other frameworks now publish CSF cross-walks; CSF's six functions provide a stable scaffold for mapping controls across regimes.

\textbf{Map IEC 62443 controls to whichever regulation applies.} IEC~62443's Security Levels and Zone/Conduit model translate cleanly into NIS2's risk-management requirements, NERC CIP's BES cyber-asset controls, and the CRA's essential cybersecurity requirements. Investing in 62443 conformance reduces friction across regimes.

\textbf{Plan for the CRA's 2027 enforcement now.} The CRA's vendor-side obligations will reshape the OT product landscape---components without secure-by-default configuration, signed firmware, and demonstrated vulnerability-handling processes will lose EU market access. Asset owners benefit by pulling those expectations forward into procurement.

\section{Related Work}
\label{sec:related_work}

OT/IT cybersecurity has accumulated a substantial survey literature over the past decade. We organize prior work by emphasis and identify the niche this survey occupies.

\subsection{Foundational Surveys of OT and ICS Threat Landscape}
The literature on industrial-control-system cybersecurity has accumulated over more than two decades. Igure et al. \cite{igure2006security} produced one of the first systematic surveys of SCADA security, cataloguing protocol- and architecture-level weaknesses before the Stuxnet era. Galloway and Hancke \cite{galloway2013introduction} surveyed industrial control networks and protocol families, providing the canonical reference for OT-protocol comparisons. C\'ardenas et al. \cite{cardenas2008research, cardenas2011attacks} formalized the cyber-physical-attack research agenda and proposed early attack taxonomies, risk assessment methods, and detection approaches that subsequent OT-IDS work built upon. Stouffer et al.'s NIST Special Publication 800-82 lineage \cite{nist80082r1, stouffer2015nist80082r2, o2024fiscal, nist80082rev3} is the principal U.S. governmental reference and has been revised to track the evolving threat landscape from Rev.~1 (2011) through Rev.~3 (2023).

More recent practitioner and academic surveys include Knapp's textbook \cite{knapp2015securing}, the canonical practitioner reference for ICS architecture and protocol-level vulnerabilities; McLaughlin et al. \cite{mclaughlin2016cps}, who surveyed the ICS cybersecurity landscape with smart-grid emphasis; and Humayed et al. \cite{humayed2017cps}, who provided a cross-domain survey of cyber-physical systems security spanning industrial, automotive, and medical settings. These surveys ground the OT threat model but predate the major incidents of 2020--2025 (Colonial Pipeline, JBS, HSE, INCONTROLLER, AcidRain, FrostyGoop, Change Healthcare, JLR) and the post-hoc financial disclosures that quantified earlier events such as NotPetya---developments that together reshaped the commercial-impact picture and established the wartime-cyber-operations precedent.

\subsection{Surveys of OT Intrusion Detection}
A second strand focuses specifically on intrusion detection in industrial environments. Vinayakumar et al.\ \cite{vinayakumar2019deep} benchmarked deep-learning intrusion-detection methods; Mirsky et al.\ \cite{mirsky2018kitsune} introduced the Kitsune ensemble-of-autoencoders approach, a lightweight unsupervised technique subsequently applied to OT and IoT settings. Recent reviews of AI/ML for ICS \cite{sarker2024ai} consolidate model-architecture surveys, and dataset reviews catalog SWaT \cite{goh2016swat}, WADI \cite{ahmed2017wadi}, HAI \cite{shin2020hai}, and Edge-IIoTset. These works are exhaustive on detection algorithms but largely silent on architectural defenses (Zero Trust, segmentation, blockchain), the broader incident-response toolchain, and commercial impact.

\subsection{IT--OT Convergence and Industry 4.0 Security}
Negi et al.\ \cite{negi2024towards}, Zaid et al.\ \cite{zaid2024emerging}, and George \cite{george2024impact} survey the cybersecurity implications of Industry~4.0 and IT--OT convergence at varying depths, emphasizing IIoT proliferation and the expanded attack surface. These reviews identify convergence as the central threat-amplifier but do not consolidate the historical incident record or quantify commercial impact.

\subsection{Sectoral and Regulatory Surveys}
Sector-specific surveys cover healthcare OT \cite{claroty2024healthcare}, energy and smart grid security \cite{waqas2024smart}, and cross-sector compliance regimes \cite{bellamkonda2020cybersecurity, fortunato2020risk}. Regulatory-focused work \cite{lanz2024updated, hatinen2024evolution, leander2019applicability} catalogs NIST CSF, NIST~SP~800-82, and IEC~62443 evolution. These works provide deep coverage of their sector or framework but are narrower in scope than a cross-sector survey.

\subsection{Incident Analyses}
The third strand consists of focused incident analyses---Stuxnet \cite{Langner, kushner2013real, farwell2011stuxnet}, Industroyer \cite{kozak2023industroyer}, Triton/Trisis \cite{mekdad2021threat, trisis_drago}, and Colonial Pipeline \cite{guardian2021colonial, cisa2023colonial}---which provide the primary-source material on which threat-landscape surveys depend. Industry threat-intelligence reports (Dragos OT Cyber Year in Review, Claroty CPS/OT survey, Fortinet \emph{State of OT Cybersecurity}, IBM X-Force Threat Intelligence Index, Mandiant M-Trends) provide annual, telemetry-grounded updates that academic surveys often lag.

\subsection{Positioning of This Survey}
Table~\ref{tab:related_work} summarizes how this survey relates to representative prior work. The contribution of this survey is to integrate four perspectives that prior surveys treat in isolation: (i)~a vector taxonomy that explicitly bridges IT-side and OT-side techniques; (ii)~a comparative tools review with an explicit gap analysis tied to the surveyed shortcomings; (iii)~a consolidated historical incident record with quantitative commercial-impact analysis; and (iv)~a sectoral deep-dive into healthcare as a worked example of IT--OT convergence under high-consequence conditions. We treat regulatory frameworks as a cross-cutting governance layer rather than the primary lens. We do not claim novel detection algorithms, architectural designs, or measured benchmarks; the contribution is synthesis and source-traceable consolidation suitable as an entry point for both practitioners and researchers.

\begin{table}[htbp]
\centering
\footnotesize
\caption{Coverage of representative prior surveys versus this work. \checkmark\ = covered in depth; \,$\circ$\,= touched briefly; --- = not covered.}
\label{tab:related_work}
\renewcommand{\arraystretch}{1.2}
\begin{tabular}{|p{5.0cm}|c|c|c|c|c|}
\hline
\textbf{Reference / scope} & \textbf{Vectors} & \textbf{Tools} & \textbf{Gaps} & \textbf{History} & \textbf{Commercial impact} \\ \hline
Igure et al.\ \cite{igure2006security} (SCADA security, foundational) & \checkmark & $\circ$ & $\circ$ & --- & --- \\ \hline
Galloway \& Hancke \cite{galloway2013introduction} (ICS networks and protocols) & $\circ$ & --- & --- & --- & --- \\ \hline
C\'ardenas et al.\ \cite{cardenas2008research, cardenas2011attacks} (CPS-attack taxonomy) & \checkmark & $\circ$ & $\circ$ & --- & --- \\ \hline
Stouffer et al.\ NIST 800-82 R1--R3 \cite{nist80082r1, stouffer2015nist80082r2, nist80082rev3} & $\circ$ & \checkmark & $\circ$ & --- & --- \\ \hline
McLaughlin et al.\ \cite{mclaughlin2016cps}; Humayed et al.\ \cite{humayed2017cps} (CPS landscape) & \checkmark & $\circ$ & $\circ$ & --- & --- \\ \hline
Knapp \cite{knapp2015securing} (ICS architecture, vulnerabilities) & \checkmark & $\circ$ & --- & $\circ$ & --- \\ \hline
Vinayakumar et al.\ \cite{vinayakumar2019deep}; Mirsky et al.\ \cite{mirsky2018kitsune} (DL IDS) & --- & \checkmark & $\circ$ & --- & --- \\ \hline
Sarker \cite{sarker2024ai} (AI in cybersecurity) & $\circ$ & \checkmark & $\circ$ & --- & --- \\ \hline
Negi et al.\ \cite{negi2024towards}; George \cite{george2024impact} (IT--OT / Industry~4.0) & \checkmark & $\circ$ & $\circ$ & $\circ$ & --- \\ \hline
Bellamkonda \cite{bellamkonda2020cybersecurity}; Lanz \cite{lanz2024updated} (regulatory) & --- & $\circ$ & --- & --- & --- \\ \hline
Claroty \cite{claroty2024healthcare}; IBM X-Force \cite{ibm2025xforce} (sectoral, healthcare) & $\circ$ & \checkmark & $\circ$ & $\circ$ & \checkmark \\ \hline
\textbf{This survey} & \checkmark & \checkmark & \checkmark & \checkmark & \checkmark \\ \hline
\end{tabular}
\end{table}

\section{Conclusion and Open Problems}
\label{sec:conclusion}

Operational Technology cybersecurity has shifted, in a single decade, from a niche concern of process engineers to a measurable component of national economic risk. This survey assembled the field along four complementary axes, answering the four research questions posed in Section~\ref{sec:introduction}.

\textbf{On vectors (RQ1).} Most documented OT compromises begin in IT and pivot inward through a small set of well-understood techniques: phishing, exposed remote access, vulnerability exploitation in internet-facing services, supply-chain compromise, and removable media. Once an attacker reaches the IT--OT boundary, OT-side propagation exploits insecure protocols, weak authentication on legacy controllers, unsigned firmware, infrequent patching, and limited audit. Section~\ref{sec:ga_sbo} mapped these IT-side and OT-side vectors into a single end-to-end taxonomy; the highest-consequence incidents have all required deep, vendor-specific OT knowledge, raising the bar simultaneously on attacker tradecraft and defender visibility.

\textbf{On tools (RQ2).} Defensive tooling now spans signature- and rule-based IDS, AI/ML-driven anomaly detection, Zero Trust Architecture, blockchain-based event logging, digital twins, and OT-aware SOC tooling. AI/ML techniques outperform signature-based detection on novel threats, particularly when paired with signature-based detection in hybrid pipelines. Architectural defenses---Zero Trust, micro-segmentation, unidirectional gateways---reduce blast radius and lateral propagation, the dominant cost driver in major incidents. Standards-aligned governance (NIST CSF, NIST~SP~800-82, IEC~62443, NIS2/DORA, IEC~80001-1) provides a consistent control vocabulary across regulatory regimes.

\textbf{On gaps (RQ3).} Section~\ref{subsec:gaps} identified seven gaps where investment yields the highest return: OT-aware patch management, labelled OT attack datasets, ML drift and explainability for safety-critical alerts, authentication retrofit for legacy protocols, IoMT and healthcare OT segmentation, consistent cross-framework resilience metrics, and supply-chain attestation for firmware and software. Each is an active research direction with real-world operational pull.

\textbf{On commercial effects (RQ4).} Section~\ref{sec:history_impact} consolidated 69 cross-validated incidents from Stuxnet (2010) through Jaguar Land Rover (2025), with the largest commercial impacts disclosed by Change Healthcare (\textasciitilde\$2.46\,B total; \cite{changehc2024}), Ascension Health (\$1.8\,B FY24 operating loss to which the cyber-incident response was a significant contributor; \cite{ascension2024}), JLR (£1.9\,B UK economy modelled by the UK Cyber Monitoring Centre; \cite{ cmc_jlr2025}), the NotPetya cascade (\textasciitilde\$10\,B globally per UK NCSC; \cite{notpetya2018wh}), CDK Global (\textasciitilde\$1.02\,B AEG-modelled dealer losses; \cite{cdkglobal2024, aeg2024cdk}), and Merck (\$870\,M from NotPetya). The dominant cost driver is business interruption, not data exfiltration. August 2025 Dragos--Marsh McLennan modelling of a severe-but-plausible 1-in-250-year tail-event scenario places global OT loss exposure at approximately US\$329.5\,B, with US\$172.4\,B attributed to OT-related business interruption; the same study estimates average annual OT-related cyber risk at approximately US\$31.1\,B \cite{dragos2025risk}. Concrete per-incident downtime costs vary dramatically by sector: Colonial Pipeline industry-estimated daily losses exceeded \$25\,M during an approximately 5-day pipeline outage \cite{cisa2023colonial}; the UK Cyber Monitoring Centre modelled JLR's UK-manufacturing losses at approximately £108\,M/week \cite{ cmc_jlr2025}. Energy and utilities, manufacturing, and transportation absorb the largest sectoral share by incident count; healthcare absorbs the highest per-incident cost \cite{ibm2025xforce}.

\subsection{Open Problems}
We highlight three cross-cutting open problems that emerge consistently across the gap analysis and the historical record.

\textbf{Trustworthy AI for OT.} The deployment gap for ML-based OT detection is no longer about accuracy on testbed datasets---hybrid pipelines routinely report 95\%+ accuracy on standard benchmarks (Table~\ref{tab:performance_comparison} and \cite{vinayakumar2019deep, DBLP:journals/corr/abs-1807-07282}). The gap is about \emph{trust under operational shift}: drift detection, calibrated false-positive rates, operator-readable explanations, and assurance that ML alerts are safe to act on automatically. Closing this gap will require methodological advances in continual learning, conformal prediction or other uncertainty quantification techniques, and human-factors research on operator interaction with ML alerts in safety-critical settings.

\textbf{Authenticated retrofits for legacy protocols.} Modbus, EtherNet/IP, and unsecured DNP3 dominate the installed base and will continue to do so for the 15--25-year lifecycle of currently deployed equipment. Engineering- and standards-grade work on in-line authenticators, proxy gateways, and protocol-aware firewalls that retrofit cryptographic authentication without controller replacement is essential and underrepresented in the literature.

\textbf{Cross-framework resilience metrics tied to operational telemetry.} Multiple resilience-quantification frameworks coexist (NIST SP~800-160 Vol.~2, DoD CREM, MITRE Cyber Resiliency Engineering, IEC~62443 SL, NERC CIP), each with its own vocabulary and emphasis. Practitioners need a small, instrumentable set of resilience metrics that can be derived from existing OT telemetry, that map cleanly to all major frameworks, and that connect directly to operational and financial KPIs (Operational Recovery Index, downtime, throughput). Achieving this requires both methodological convergence among the frameworks and engineering work on telemetry instrumentation in deployed OT.

\subsection{Closing}
The historical record shows that adversary capability has consolidated faster than defender capability. Bespoke nation-state operations from a decade ago have given way to reusable toolkits and OT-aware ransomware, while the lower end of the spectrum has democratized to opportunistic exploitation of internet-exposed PLCs running default credentials. The defender response is also visible---AI/ML detection, Zero Trust, OT-aware SOCs, sector-specific regulation---but uneven across sectors and lagging in those with the longest equipment lifecycles. On the evidence assembled here, defensive posture appears most consistent with strong outcomes when it combines AI/ML-driven detection at the IT--OT boundary, Zero-Trust segmentation to constrain blast radius, OT-aware incident-response automation to reduce time-to-contain, and standards-aligned governance to make these investments durable; rigorous cost-effectiveness analysis across sectors and incident types remains an open research direction. Lessons from past incidents, combined with the seven gaps identified in Section~\ref{subsec:gaps}, provide a concrete roadmap for both research and practice.


\section*{CRediT authorship contribution statement}
\textbf{Harsh Vardhan:} Conceptualization, Methodology, Investigation, Data curation, Writing -- original draft, Writing -- review \& editing, Visualization, Project administration.
\textbf{Sarthak Kapoor:} Investigation, Data curation, Formal analysis, Writing -- original draft, Visualization.
\textbf{Sumit Kumar:} Investigation, Data curation, Writing -- review \& editing.

\textbf{Daniel Balasubramanian:} Conceptualization, Supervision, Writing -- review \& editing.
\textbf{Sandeep Neema:} Supervision and Guidance.

\section*{Declaration of competing interest}
The authors declare that they have no known competing financial interests or personal relationships that could have appeared to influence the work reported in this paper.

\section*{Funding}

This work was supported partially by DARPA CASTLE project [grant number SFP\_302314].

\section*{Data availability}
No new data were created or analyzed in this study. All incident records and figures are derived from publicly available sources cited in the article.

\section*{Declaration of generative AI and AI-assisted technologies in the manuscript
preparation process}
During the preparation of this work the author(s) used Claude in order to
check for issues relating to spelling, grammar, and confusing language. After using
this tool/service, the author(s) reviewed and edited the content as needed and take(s)
full responsibility for the content of the published article.
\bibliographystyle{elsarticle-num} 
\bibliography{publications}

\end{document}